\documentclass[journal]{IEEEtran}
\IEEEoverridecommandlockouts
\usepackage{cite}
\usepackage{amsmath,amssymb,amsfonts}
\usepackage{algorithmic}
\usepackage{graphicx}
\usepackage{textcomp}
\usepackage{xcolor}

\usepackage{orcidlink}
\usepackage{hyperref}
\usepackage[switch]{lineno}

\def\BibTeX{{\rm B\kern-.05em{\sc i\kern-.025em b}\kern-.08em
    T\kern-.1667em\lower.7ex\hbox{E}\kern-.125emX}}
\begin{document}

\title{Interfacing Superconductor and Semiconductor Digital Electronics\\
}

\author{Yerzhan~Mustafa and~Selçuk~Köse
\thanks{Y. Mustafa and S. Köse are with the Department
of Electrical and Computer Engineering, University of Rochester, Rochester,
NY, 14627, USA. E-mails: (yerzhan.mustafa@rochester.edu,  selcuk.kose@rochester.edu).}
\thanks{This work is supported in part by the National Science Foundation Expeditions Award under Grant CCF-2124453 and SHF Award under Grant CCF-2308863, and Department of Energy EXPRESS program under Grant DE-SC0024198.}
}

\maketitle

\begin{abstract}
Interface circuits are the key components that enable the hybrid integration of superconductor and semiconductor digital electronics. 
The design requirements of superconductor-semiconductor interface circuits vary depending on the application, such as high-performance classical computing, superconducting quantum computing, and digital signal processing. 
In this survey, various interface circuits are categorized based on the working principle and structure. 
The superconducting output drivers are explored, which are capable of converting and amplifying, \textit{e.g.}, single flux quantum (SFQ) voltage pulses, to  voltage levels that semiconductor circuits can process. 
Several trade-offs between circuit- and system-level design parameters are examined. 
Accordingly, parameters such as the data rate, output voltage, power dissipation, layout area, thermal/heat load of cryogenic cables, and bit-error rate are considered. 

\end{abstract}

\section{Introduction}

Superconductor digital electronics such as single flux quantum (SFQ) logic is a promising candidate for beyond-CMOS (complementary metal–oxide–semiconductor) technology due to the high-speed operation of tens to hundreds of GHz and low energy consumption of 10$^{-19}$ J per switching activity \cite{likharev1991rsfq,holmes2013energy,przybysz2015superconductor,braginski2019superconductor,krylov2024single}. 
SFQ circuits comprise a wide variety of logic families such as rapid SFQ (RSFQ) \cite{likharev1991rsfq}, energy-efficient RSFQ (ERSFQ) \cite{kirichenko2011zero}, energy-efficient SFQ (eSFQ) \cite{mukhanov2011energy}, reciprocal quantum logic (RQL) \cite{herr2011ultra}, adiabatic quantum flux parametron (AQFP) \cite{takeuchi2013adiabatic}, pulse conserving logic (PCL) \cite{herr2023superconducting}, and superconducting sustainable ballistic fluxon (SSBF) \cite{lupo2025digital}. All of these logic families use Josephson junctions (JJs) as a switching element and operate at cryogenic temperatures (\textit{e.g.}, 4 K). 

Despite delivering significant performance gains, superconductor digital electronics are not expected to fully replace advanced semiconductor (\textit{e.g.}, CMOS) technology. 
A hybrid approach, which leverages the key advantages of both superconductor and semiconductor technologies, is beneficial. Early applications of superconductor digital electronics include microprocessors \cite{ando2016design,ishida2018towards,ayala2021mana,kashima2021microprocessor,nagaoka2022microprocessor,tanaka2023execution}, accelerators \cite{herr2024data,kundu2025system}, digital radio frequency (RF) receivers \cite{wikborg1999rsfq,kirichenko2005superconductor,gupta2007digital,mukhanov2008superconductor}, and control and readout circuits for superconducting quantum computers \cite{mcdermott2014accurate,patel2017phonon,mcdermott2018quantum,mukhanov2019scalable,jokar2022digiq,liu2023single,barbosa2024rsfq,di2023discriminating,di2024fast,di2025control,bernhardt2025quantum}. Semiconductor electronics is typically used to complement certain limitations of superconductor circuits such as low-density memory \cite{nagasawa1995380,nagasawa2007yield,konno2017fully,van201364,hironaka2020demonstration}.

The hybrid integration of superconductor and semiconductor electronics requires special interface circuits to convert signaling between two different technologies. 
Semiconductor circuits (\textit{e.g.}, CMOS) encode logical `1' and `0' into constant (DC) high and low voltage levels, respectively, in the order of hundreds of mV to a few volts, depending on the fabrication node. 
Superconductor circuits such as SFQ logic encode logical `1' and `0' into the presence and absence of voltage pulses, respectively. These voltage pulses are typically about 1 mV in amplitude and approximately 2 ps in duration and have a constant value equal to one magnetic flux quantum when integrated over time ($\Phi_0\approx 2.07\times10^{-15}$ Wb) \cite{likharev1991rsfq}. 
A particular challenge is to convert and amplify SFQ pulses to appropriate voltage signals that CMOS circuits can process. 
The low driving capability of JJs introduces certain trade-offs in amplification gain, power dissipation, layout area, and other parameters of superconductor-semiconductor interface circuits.

In this paper, various interface circuits, which are used between superconductor and semiconductor digital electronics, are explored. 
First, existing applications of superconductor digital electronics are discussed, and the need for interface circuits is detailed in Section \ref{section:applications}. 
A taxonomy of superconductor-semiconductor interface circuits is presented in Section \ref{section:taxonomy} with a detailed discussion of the working principle and structure of each circuit type. 
The implications of different fabrication technologies on the performance of interface circuits are analyzed in Section \ref{section:fabrication}. 
A comparison of superconducting output drivers, a subset of superconductor-semiconductor interface circuits, is provided based on circuit- and system-level criteria in Section \ref{section:comparison}. 
The conclusions are drawn in Section \ref{conclusion}.

\section{The Need for Superconductor-Semiconductor Interface Circuits}\label{section:applications}

The superconductor-semiconductor interface circuits are essential building blocks in establishing communication between various cryogenic (superconducting) technologies and the room-temperature electronics. 
The following is a list of emerging applications that utilize superconductor digital electronics and require a specialized interface to semiconductor circuits. 

\textbf{High-performance computing (HPC)}: 
Superconductor digital electronics (\textit{e.g.}, SFQ logic) can operate at extremely high switching frequency of tens to hundreds of GHz while consuming 10$^{-19}$ J energy per switching activity \cite{likharev1991rsfq,holmes2013energy,przybysz2015superconductor,braginski2019superconductor,krylov2024single}. 
These characteristics make this technology a promising candidate for beyond-CMOS technology, particularly, for HPC such as data centers and cloud computing \cite{holmes2013energy,przybysz2015superconductor,braginski2019superconductor,krylov2024single}. 
Several 8-bit microprocessors have been designed using SFQ logic \cite{ando2016design,ishida2018towards,kashima2021microprocessor,nagaoka2022microprocessor,tanaka2023execution}.
For example, the microprocessor in~\cite{nagaoka2022microprocessor} is capable of operating at 57.2 GHz with a power consumption of 11.2 mW. 
An SFQ-based microprocessor needs a cryogenic memory to store and load various instructions and data. 
Various types of cryogenic memory are available in the literature, as reviewed in \cite{alam2023review}. 
A hybrid approach, which uses semiconductor-based (\textit{e.g.}, CMOS) memory, offers high storage capacity/density and fabrication maturity as compared to other cryogenic memories \cite{alam2023review}. 
This type of memory is known as Josephson-CMOS hybrid memory \cite{konno2017fully,van201364,hironaka2020demonstration}. 
Therefore, interface circuits are needed to convert signals between superconducting microprocessor and semiconductor memory. 
A recent spin-out of IMEC, Snowcap Compute, Inc., is working on building artificial intelligence computing chips using superconductor digital electronics \cite{herr2024data,snowcap_compute}.

\textbf{Quantum computing}: To realize a quantum computer that can compete with modern CMOS based computing devices, a large number of qubits are needed. 
One promising approach to scaling superconducting quantum computers is placing the qubit control and readout circuitry in close proximity to the qubits (\textit{i.e.}, within the same temperature zone) to reduce the thermal load from cryogenic interconnects. 
SFQ technology is an emerging approach to the control and readout of superconducting qubits because of its ultra-low-power dissipation \cite{mcdermott2014accurate,patel2017phonon,mcdermott2018quantum,mukhanov2019scalable,jokar2022digiq,liu2023single,barbosa2024rsfq}.
Companies such as SEEQC and Atlantic Quantum (recently acquired by Google Quantum AI \cite{google_quantum_ai}) are working on developing quantum computers with superconductor digital circuits to enable large-scale control and readout \cite{di2023discriminating,di2024fast,di2025control,bernhardt2025quantum,kannan2024managing}. 

Qubits are prone to errors (decoherence) due to sensitivity to the environment. 
To overcome this challenge, quantum error correction (QEC) is needed. 
Fast, real-time correction of qubits can be achieved using graphics processing units (GPUs) due to their highly parallelizable processing capabilities. 
GPUs, which are made with semiconductor technology, can accelerate the decoding of error syndromes and determine the required corrections. 
High-speed superconductor-semiconductor interface circuits can therefore facilitate fast QEC. 
For example, SEEQC in collaboration with NVIDIA develops the first fully digital quantum-classical interface for ultra-low latency QEC \cite{seeqc_nvidia,seeqc_nvidia2}.

\textbf{Digital signal processing}: 
A software-defined radio is a communication system in which instead of analog hardware, broadband RF signals are converted into digital form \cite{mitola2000software}. 
Due to the high frequency operation of SFQ logic (\textit{i.e.}, tens of GHz), superconductor digital RF receivers are capable of direct digitization of RF signals in the range of kHz to GHz \cite{wikborg1999rsfq,kirichenko2005superconductor,gupta2007digital,mukhanov2008superconductor}. 
For example, HYPRES, Inc. offers a commercially available digital RF receiver \cite{hypres_RF}. 
This system uses 17-channel interface amplifiers, which enable communication between superconductor (\textit{i.e.}, SFQ) and semiconductor (\textit{i.e.}, field-programmable gate array (FPGA)) circuits \cite{hypres_RF}.

\textbf{Single-photon detectors}: Superconducting single-photon detectors include superconductor nanostrip photon detector (SNSPD) \cite{gol2001picosecond,natarajan2012superconducting} and transition edge sensor (TES) \cite{hummatov2023fast,gottardi2021review,ullom2015review}. Note that SNSPD nomenclature is based on International Electrotechnical Commission standard, which redefines commonly used `superconducting nanowire single-photon detector' term \cite{SNSPD_IEC}.
These detectors can provide higher detection efficiency, lower dark counts, and faster response times as compared to semiconductor-based detectors \cite{natarajan2012superconducting,gottardi2021review,ullom2015review}. Nevertheless, single-photon detectors should operate at cryogenic temperatures and, hence, require an interface circuitry to send the data to room-temperature electronics. 
For example, superconductor electronic circuits are used to readout SNSPD \cite{miyajima2017timing,miki2018superconducting,shelly2017modelling,komissarov2024modified,sobolewski2025electrical,zheng2019characterize,castellani2024nanocryotron} and TES \cite{bozbey2009single,leman2023integrated,beyer2003performance,durkin2019demonstration,kiviranta2021two,kiviranta2025two}.

\textbf{Neuromorphic computing}: 
Superconductor electronics can be used to implement neuromorphic computing systems at the hardware-level. 
JJs are basic building blocks of superconductor electronic circuits \cite{likharev1991rsfq}. 
The spiking nature of brain neurons could be mimicked with JJs, which produce voltage pulses during switching \cite{schneider2022supermind}. 
Moreover, a near lossless transmission of these voltage pulses over superconducting transmission lines enables fast and energy-efficient operation of neuromorphic circuits \cite{schneider2022supermind}. 
Neuromorphic computing hardware types that are implemented with superconductor electronic circuits and devices have been reviewed in \cite{schneider2022supermind,islam2023review}.
Neuromorphic computing requires memory elements to store synapse weights \cite{schneider2022supermind}. 
While novel superconductor-based memory is an ongoing research, the hybrid approach that uses mature, high-capacity semiconductor memory is preferred similar to HPC systems \cite{alam2023review}. 
Additionally, superconductor-semiconductor interface circuits are essential components for testing the performance of novel superconducting neuromorphic circuits and architectures as, \textit{e.g.}, implemented in \cite{karamuftuoglu2023jj,razmkhah2024hybrid}.

The aforementioned applications use superconductor-semiconductor interface circuits under various design requirements and limitations such as data rate, gain (amplification), operating temperature (\textit{e.g.}, 20 mK, 4 K, 50 K, etc.), power dissipation, and fabrication technology. 
As a result, various types of interface circuits with different characteristics have been proposed in the literature. 
Therefore, it is important to categorize and identify the unique advantages and  drawbacks of each type of interface circuit.

\section{Taxonomy of Superconductor-Semiconductor Interface Circuits}\label{section:taxonomy}

\begin{figure*}
  \centering
    \includegraphics[width=1\textwidth]{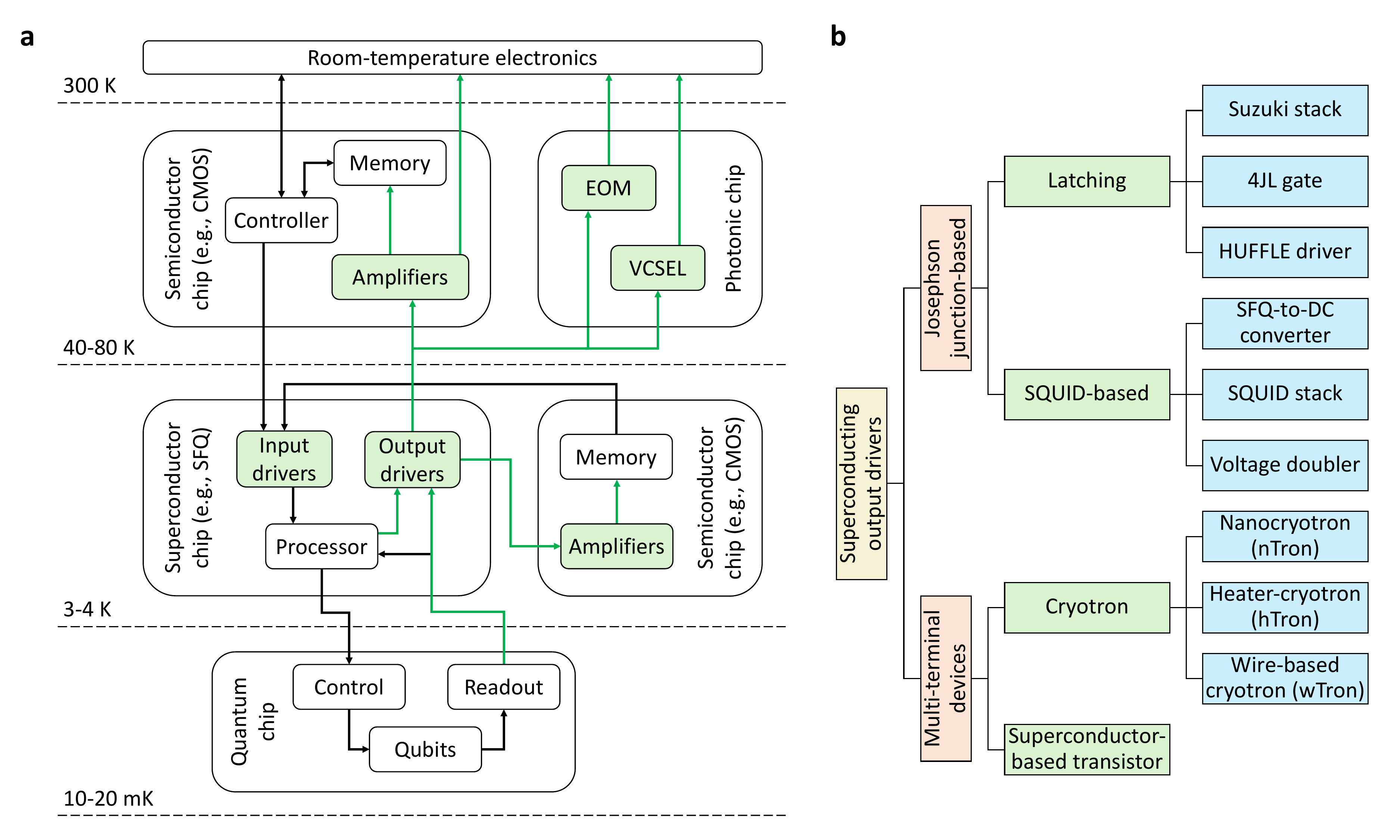}
    \caption{\textbf{a}, Block diagram of the cryostat with superconductor, semiconductor, photonic, and quantum chips. The superconductor-semiconductor interface circuits are highlighted in green boxes. Digital output data links are shown with green arrows. \textbf{b}, Taxonomy of the superconducting output drivers.}
    \label{fig:Cryostat_block_diagram_and_taxonomy}
\end{figure*}

Digital data links with superconductor-semiconductor interface circuits typically consist of two stages. The first stage is a superconducting output driver that receives a signal from superconductor digital electronics (\textit{e.g.}, SFQ pulse), converts, and amplifies it to a DC signal in the range of hundreds of \textmu V to hundreds of mV. The second stage is a semiconductor amplifier that further boosts the DC signal up to the level of CMOS technology (around 1 V) \cite{gupta2019digital}. 
Alternatively, a photonic circuit can be used as the second stage such as an electro-optic modulator (EOM) \cite{sakai2020proposal,youssefi2021cryogenic,de2021attojoule,pintus2024cryogenic,yin2021electronic,shen2024photonic,yin2026fully} or vertical cavity surface emitting lasers (VCSEL) \cite{mukhanov2013development,wu2021vcsel,wu2024cryo,namvar2024improving,namvar2025thermal}. 
A block diagram of the cryostat housing superconductor, semiconductor, photonic, and quantum chips is depicted in Fig. \ref{fig:Cryostat_block_diagram_and_taxonomy}(a).
The key components of the superconductor-semiconductor interface are highlighted in green in Fig. \ref{fig:Cryostat_block_diagram_and_taxonomy}(a). 
It should be noted that the input drivers (see Fig. \ref{fig:Cryostat_block_diagram_and_taxonomy}(a)) are also part of the superconductor-semiconductor interface circuits.
Due to the straightforward nature of converting DC voltage signals from semiconductor circuits to SFQ pulses, input-side interfaces such as DC-to-SFQ converters
\cite{kaplunenko1989experimental,sunysb_rsfq_cell_library,supertools_rsfq_cell_library,li2024cmos} are commonly reused across many designs. 
In contrast, the output stage presents substantially greater design difficulty and admits a wider range of circuit implementations. 
Therefore, the focus of this work is placed on the output drivers.

A taxonomy of superconducting output drivers is shown in Fig. \ref{fig:Cryostat_block_diagram_and_taxonomy}(b).
In this classification, output drivers are grouped into two major categories: 1) JJ-based and 2) multi-terminal devices.

\subsection{JJ-based output drivers}\label{subsection:JJ_based_drivers}

JJ-based output drivers use JJs as a switching element. 
A JJ is a two-terminal device that consists of two superconductor materials separated by a weak link (e.g., Nb/AlO$_x$/Nb). 
JJs are commonly used in SFQ logic and superconducting quantum circuits, and have multiple established fabrication processes (\textit{e.g.}, MIT Lincoln Lab SFQ5ee \cite{tolpygo2016advanced}, SEEQC SFQ-C5SL \cite{yohannes2023high}, ADP2 \cite{nagasawa2014nb}, SIMIT Nb03P \cite{ying2021development}, and FLUXONICS CJ2 \cite{kunert2024advanced}). 
Depending on the type of JJ damping, JJ-based output drivers can be grouped into two classes, as depicted in (Fig. \ref{fig:Cryostat_block_diagram_and_taxonomy}(b)). 
The first class consists of latching drivers, which employ underdamped junctions characterized by a Stewart-McCumber parameter $\beta_c > 1$ \cite{braginski2019superconductor}.
The second class consists of SQUID (superconducting quantum interference device)-based drivers, which rely on overdamped  or critically damped junctions, respectively, with $\beta_c < 1$ and $\beta_c=1$.

\subsubsection{\textbf{Latching drivers}}

In latching drivers, the underdamped JJs exhibit hysteretic behavior. Once the current flowing through these types of JJs exceeds the critical value, the junction switches from superconducting state to resistive state. 
The JJ is latched in the resistive state until its current is reduced below a specific threshold value (known as return current). 
The following are examples of latching driver circuits.

\begin{figure*}
  \centering
    \includegraphics[width=0.8\textwidth]{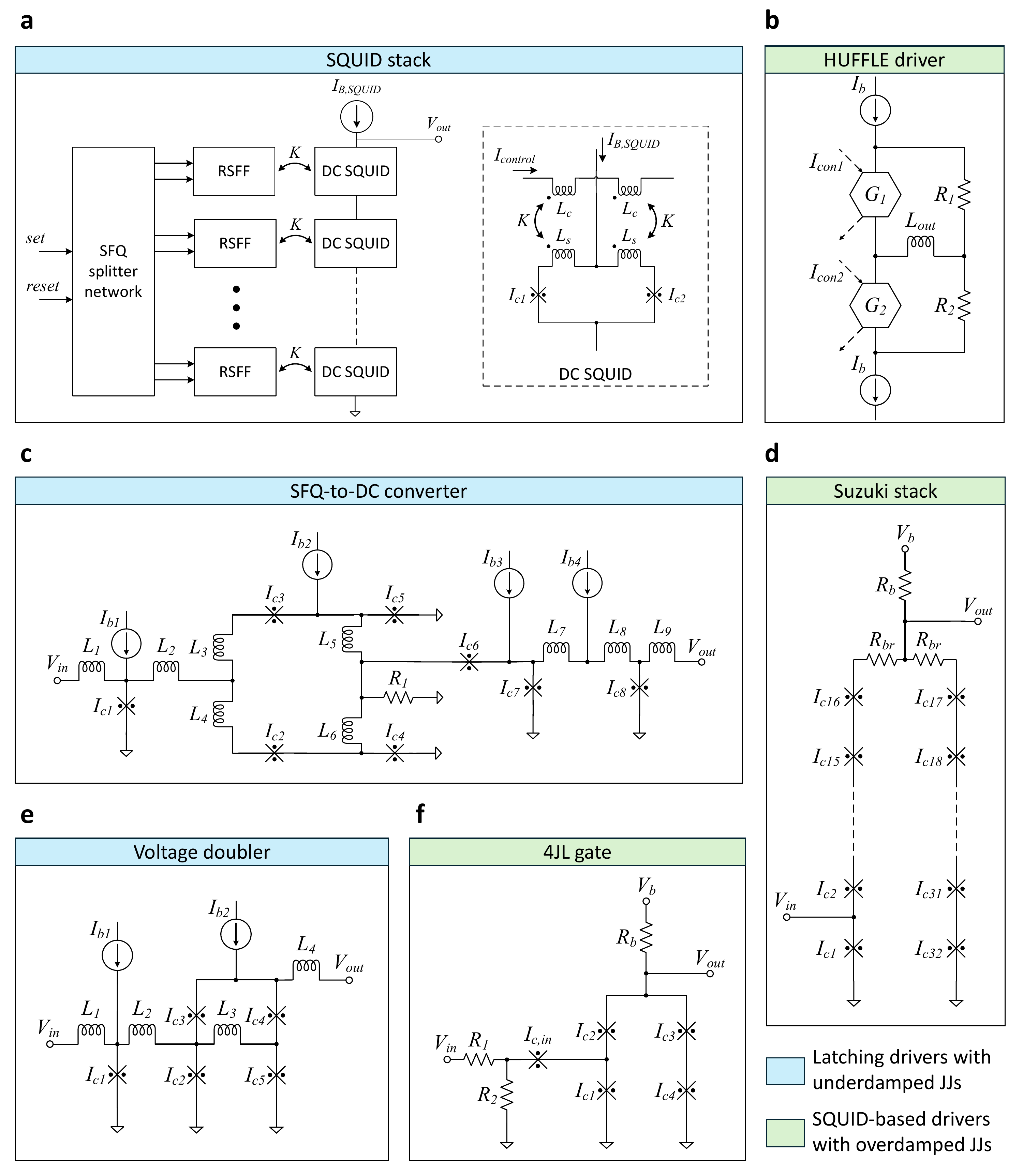}
    \caption{Schematics of JJ-based output drivers. \textbf{a}, SQUID stack with RSFFs \cite{hashimoto2007implementation}. The inset shows the schematic of DC SQUID cell. \textbf{b}, HUFFLE driver \cite{hebard1979dc}. \textbf{c}, SFQ-to-DC converter based on TFF \cite{likharev1991rsfq,supertools_rsfq_cell_library}. \textbf{d}, Suzuki stack with 16-JJ configuration \cite{ortlepp2013design}. \textbf{e}, Voltage doubler \cite{ortlepp2009superconductor}. \textbf{f}, 4JL gate \cite{konno2017fully}. Shunt resistors and parasitic inductances of JJs are not shown. A novel symbol of JJ is used that is standardized by International Electrotechnical Commission \cite{new_JJ_symbol}.}
    \label{fig:JJ_based_drivers_schematics}
\end{figure*}

\begin{figure*}
  \centering
    \includegraphics[width=1\textwidth]{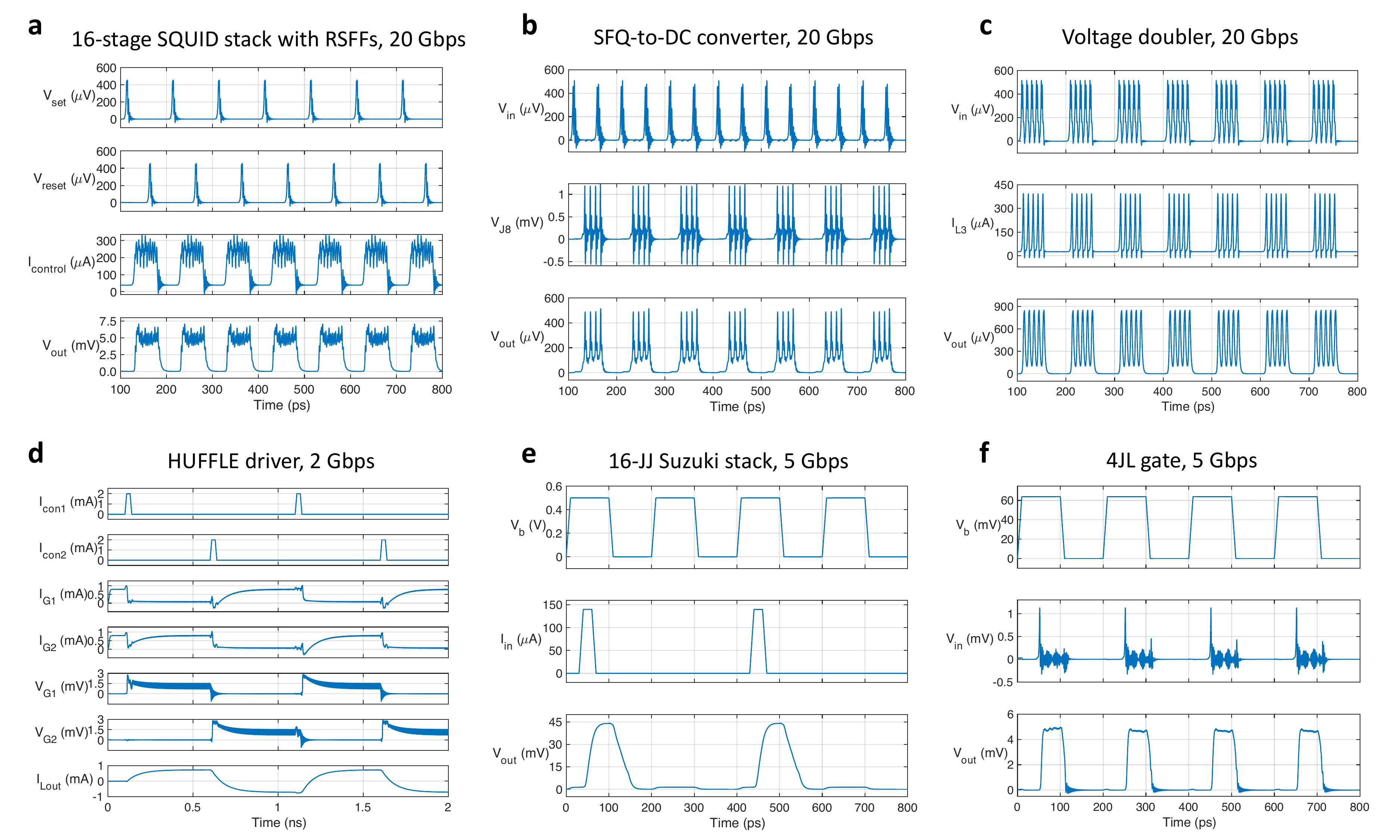}
    \caption{Simulation of JJ-based output drivers. Simulated in JoSIM \cite{delport2019josim} using MIT Lincoln Lab SFQ5ee process with the critical current density of 100~\textmu A/\textmu m$^2$ \cite{tolpygo2016advanced}. \textbf{a}, 16-stage SQUID stack with RSFFs, 
    \textbf{b}, SFQ-to-DC converter, and 
    \textbf{c}, voltage doubler are operating at 20 Gbps with 50 $\Omega$ and 100 pH load. The input signal of voltage doubler is modeled as a 100 GHz SFQ pulse train \cite{ortlepp2009superconductor}. 
    \textbf{d}, HUFFLE driver operating at 2 Gbps with $L_{out}$ = 100 pH. 
    \textbf{e}, 16-JJ Suzuki stack is operating at 5 Gbps with 50 $\Omega$, 100 pH, and 180 fF load. 
    \textbf{f}, 4JL gate circuit is operating at 5 Gbps with 50 $\Omega$ and 100 pH load. The circuit parameters are similar to designs presented in \cite{mustafa2025spam,supertools_rsfq_cell_library,kaplunenko1991single,hebard1979dc,ortlepp2013design,konno2017fully}.}
    \label{fig:JJ_based_drivers_waveforms}
\end{figure*}

\textbf{Four-junction logic (4JL) gate}: A 4JL gate was initially proposed in 1980 by Takada \textit{et al.} \cite{takada1980current,nakagawa1982operating} as a logic gate (AND, OR, and NOT) for superconducting latching logic. 
While latching logic was largely displaced by SFQ technology in the early 1990s due to the higher available clock rates \cite{przybysz2015superconductor}, the 4JL gate circuit remains a practical and effective option for use as a superconducting output driver. 
The schematic of a 4JL gate is depicted in Fig. \ref{fig:JJ_based_drivers_schematics}(f), which consists of five underdamped JJs, where four JJs come from the original logic circuit \cite{takada1980current} and one JJ is used as a buffer to prevent a back-action to the SFQ input circuits. 
The critical current of these JJs is chosen such that $I_{c1} = I_{c2} = I_{c,in} = $ 100 \textmu A and $I_{c3} = I_{c4} = $ 300 \textmu A \cite{konno2017fully,china2022high,hironaka2025josephson}. 
The operating principle of the 4JL gate is illustrated through simulations in Fig. \ref{fig:JJ_based_drivers_waveforms}(f). 
All JJs are initially biased at approximately 80\% of their critical currents, and an SFQ pulse is applied at the input terminal $V_{in}$. 
This SFQ pulse first switches the bottom left JJ with critical current $I_{c1}$ from the superconducting state to the resistive state. 
The resulting imbalance in the bias (supply) current distribution directs most of the bias current into the JJs on the right branch (with $I_{c3}$ and $I_{c4}$), effectively switching them to the resistive state. 
Finally, the top left JJ (with $I_{c2}$) and the buffer JJ (with $I_{c,in}$) transition to the resistive state. 
Once all JJs switch to the resistive state, the circuit produces an output voltage approximately twice the gap voltage 2$V_g$. 
In Fig. \ref{fig:JJ_based_drivers_waveforms}(f), an output voltage of about 5 mV is produced for MIT Lincoln Lab SFQ5ee process \cite{tolpygo2016advanced} with $V_g=2.8$ mV. 
To restore the 4JL gate to the superconducting state, the bias voltage $V_b$ is turned off once per clock cycle, generating a square-wave (AC) waveform, as shown in Fig. \ref{fig:JJ_based_drivers_waveforms}(f). 

The 4JL gate is commonly used in 4 K Josephson-CMOS hybrid memories as a pre-amplifier for Suzuki stack circuit \cite{konno2017fully,van201364,hironaka2020demonstration,hironaka2025josephson}.
Additionally, 4JL gate can be used as an output driver for digital data links from 4 K to 50 K temperature zones \cite{krause2025signal}.

\textbf{Suzuki stack}: A Suzuki stack (a.k.a. Josephson latching driver) was first proposed in 1988 by Suzuki \textit{et al.} \cite{suzuki1988josephson}. 
A schematic of a Suzuki stack circuit is depicted in Fig. \ref{fig:JJ_based_drivers_schematics}(d). 
The Suzuki stack consists of two arrays of serially connected underdamped JJs.
Each branch has a small resistor $R_{br}$ (around 3-4 $\Omega$) to avoid flux quantization \cite{ortlepp2013design}. 
In a conventional design, all JJs have an equal critical current (\textit{e.g.}, 400 \textmu A) \cite{ortlepp2013design,konno2017fully,hironaka2025josephson}. 
For an $n$ number of JJs in series, the Suzuki stack can produce an output voltage of $nV_g$. 
Depending on the application, typical values of $n$ are in the range of 4 to 24 \cite{van201364,ortlepp2013design,bhat199910,ortlepp2013high,harada2000high,hato2003high,suzuki1988josephson,suzuki1990applications,suzuki1999interface,kunert2024advanced,konno2017fully,harada2002high,przybysz1997interface,przybysz1999dewar,hironaka2025josephson,harada2004josephson,feng2003josephson,harada2003logic,hato2005output,mustafa2022optimization,mustafa2023suzuki,liu2005simulation,liu2007latency,fujiwara2010new,kojima2005parameter,harada2000multigigahertz,sandell1999high,van2002hybrid}.
The operating principle of a 16-JJ (\textit{i.e.}, $n=16$) Suzuki stack circuit is shown in Fig. \ref{fig:JJ_based_drivers_waveforms}(e). 
The input signal is a 140 \textmu A current pulse, which models the signal sent from the 4JL gate pre-amplifier \cite{ortlepp2013design}. 
Although a Suzuki stack can be driven with an SFQ pulse as was demonstrated in \cite{harada2000high,harada2002high,harada2003logic,harada2004josephson,kunert2024advanced}, a 4JL gate pre-amplifier offers improved input and output stability \cite{konno2017fully}. 
Additionally, the lowermost left JJ (with $I_{c1}$) in the Suzuki stack can be replaced with two JJs in series \cite{ortlepp2013high,konno2017fully,hironaka2025josephson}, 2-JJ (or 3-JJ) SQUID structure \cite{harada2004josephson,kojima2005parameter} for higher input sensitivity and wider bias margins.

With MIT Lincoln Lab SFQ5ee process, the 16-stage Suzuki stack can produce an output voltage of around 45 mV as shown in Fig. \ref{fig:JJ_based_drivers_waveforms}(e). 
The switching mechanism of the Suzuki stack is similar to that of the 4JL gate \cite{mustafa2023suzuki}.

Both 4JL gate and Suzuki stack typically require an AC bias voltage, which introduces several challenges including synchronization issues, coupling effects, and increased heat load of cryogenic bias cables \cite{konno2017fully,suzuki2007superconducting,klostermann1991heat}. 
The coupling effects can be mitigated by adding a top niobium (Nb) layer over the SFQ circuits, which acts as a magnetic shield \cite{konno2017fully}.
Alternatively, DC-biased Suzuki stack designs have been proposed too address these drawbacks in \cite{suzuki2007superconducting,suzuki2009possible,mustafa2024dc}. 

In a multi-channel setup with a single power supply, Suzuki stacks require a large bias resistor $R_b$ (Fig. \ref{fig:JJ_based_drivers_schematics}(d)), typically around 750 $\Omega$ \cite{ortlepp2013design,konno2017fully}. 
Consequently, most of the power is dissipated across $R_b$ \cite{ortlepp2013design}. 
To address this, a power optimization technique is proposed in \cite{mustafa2022optimization}, which employs unequal critical current values for the JJs in the left and right branches.
This approach effectively reduces power dissipation by 30-70\%. 

As noted earlier, Suzuki stack circuits (with 4JL gate pre-amplifiers) are widely used in 4 K Josephson-CMOS hybrid memories \cite{konno2017fully,van201364,hironaka2020demonstration}. 
Among the various JJ-based output drivers, a Suzuki stack is capable of producing the highest output voltages, typically ranging from tens to hundreds of millivolts \cite{van201364,ortlepp2013design,ortlepp2013high,suzuki1988josephson,mustafa2023suzuki,hironaka2025josephson}. 
A higher output voltage of Suzuki stack circuit is particularly necessary in 4 K to 4 K digital data links (\textit{e.g.}, sending the data from an SFQ chip to 4 K memory as shown in Fig. \ref{fig:Cryostat_block_diagram_and_taxonomy}(a)). 
As a result, a subsequent semiconductor amplifier stage would require a lower gain and, hence, dissipate less power, which benefits the overall cooling power budget at 4 K.
Although the operating frequency of Suzuki stack circuits is limited to a few GHz due to the latching-type of switching, this frequency is sufficient for direct interfacing with CMOS memory. 
For example, a 4 K Josephson-CMOS memory operating at 1 GHz is demonstrated in \cite{van201364}.

\textbf{Hybrid unlatching flip-flop logic element (HUFFLE) driver}:
The HUFFLE circuit was introduced in 1979 by Hebard \textit{et al.} \cite{hebard1979dc} as a DC-biased flip-flop logic element for latching logic. 
The schematic of a HUFFLE circuit is depicted in Fig. \ref{fig:JJ_based_drivers_schematics}(b). 
The circuit comprises two gates $G_1$ and $G_2$, which can be, for example, individual underdamped JJs or interferometers \cite{hebard1979dc}, along with an output inductor $L_{out}$ and two resistors, $R_1$ and $R_2$. 
The circuit includes two DC bias lines and two control currents ($I_{con1}$ and $I_{con2}$), as illustrated in Fig. \ref{fig:JJ_based_drivers_schematics}(b). 
The operating principle of the HUFFLE circuit is shown in Fig. \ref{fig:JJ_based_drivers_waveforms}(d). 
$G_1$ and $G_2$ are used as individual JJs coupled to inductive lines with $I_{con1}$ and  $I_{con2}$, respectively. 
Initially, both $G_1$ and $G_2$ are in the superconducting state, allowing bias current to flow through them. 
When $I_{con1}$ is applied, $G_1$ switches to the resistive state, diverting the bias current through $R_1$, $L_{out}$, and $G_2$. Although the current in $G_1$ remains above the return current, $G_1$ stays in the resistive state. 
Upon application of $I_{con2}$, $G_2$ switches to the resistive state, momentarily reducing the bias current through $G_1$ and causing it to return to the 
superconducting state.
After that, the bias current flows through $R_2$, $L_{out}$, and $G_1$. 

The HUFFLE circuit has been demonstrated in a ring oscillator in \cite{kotera1983ring} producing roughly 2 mV at 0.5 GHz frequency.
A comprehensive theoretical analysis of the HUFFLE circuit is presented in \cite{hatano1992performance}. 
The HUFFLE circuit has been used as an output driver in \cite{schneider1995broadband,polonsky1991new}. 
In \cite{schneider1995broadband}, $G$ is a 3-JJ interferometer, which is controlled by the SFQ-to-DC converter. 
This driver was able to produce roughly 1.5 mV with an estimated frequency of 4 GHz. 
In \cite{polonsky1991new}, $G$ is an SFQ-to-DC converter with unshunted (\textit{i.e.}, underdamped) JJs. 
This driver produced an output voltage of 2.8 mV \cite{polonsky1991new}. 
While the HUFFLE driver offers a DC-biased design without coupling and synchronization issues, it is prone to a parasitic `hang-up' (`latch-up') mode, where both $G_1$ and $G_2$ latch to the voltage state, resulting in incorrect operation \cite{hatano1992performance}. 
Therefore, this type of driver has not been commonly used in modern superconductor-based applications.

\subsubsection{\textbf{SQUID-based drivers}}
SQUID-based drivers use a DC SQUID structure \cite{clarke2006squid}, which consists of two overdamped (or critically damped) JJs in parallel, forming a superconducting loop, as shown in Fig. \ref{fig:JJ_based_drivers_schematics}(a). 
A DC SQUID produces a periodic voltage response due to an external magnetic field \cite{clarke2006squid}. 
Its period is equal to a magnetic flux quantum ($\Phi_0$), and the maximum output voltage is achieved at the integer multiples of $\Phi_0$/2. 
SQUID-based drivers leverage this effect by digitally switching the DC SQUID(s) between zero applied flux and $\Phi_0$/2 flux. 
The following are examples of SQUID-based output driver circuits.

\textbf{SQUID stack}: A SQUID stack consists of multiple DC SQUIDs that are connected in series, as shown in Fig. \ref{fig:JJ_based_drivers_schematics}(a). 
The first demonstration of a SQUID stack structure was made in 1991 by Welty \textit{et al.} \cite{welty1991series}, where 100 DC SQUIDs were fabricated using trilayer Nb/AlO$_x$/Nb JJs. 
In a series connection, a DC SQUID can produce a higher output voltage by constructively interfering the switching from each DC SQUID. 
In a SQUID stack, each DC SQUID cell is magnetically coupled to a line with the control current $I_{control}$. 
To produce the maximum output voltage, $I_{control}$ should be set such that $L_m I_{control} = \Phi_0/2$ \cite{herr2010high}. 
$L_m$ is the mutual inductance that can be expressed as $L_m = 2K\sqrt{L_cL_s}$, where $L_c$ and $L_s$ are the inductances, and $K$ is the coupling coefficient, as shown in Fig. \ref{fig:JJ_based_drivers_schematics}(a). 

Several SFQ circuits have been proposed in the literature to generate the required $I_{control}$. 
For example, Fig. \ref{fig:JJ_based_drivers_schematics}(a) illustrates a reset-set flip-flop (RSFF)-based approach in which the storage loop of the RSFF is magnetically coupled to the DC SQUID cell. 
The RSFF-based SQUID stack designs have been reported in \cite{koch1999nrz,dubash2000system,hashimoto2007implementation,hashimoto2009measurement,terai2009readout,zhao2025high,zhao2025high_magnetometer,mustafa2025spam}. 
The operating principle of a 16-stage SQUID stack employing RSFFs is illustrated in Fig. \ref{fig:JJ_based_drivers_waveforms}(a). 
A $set$ signal introduces a fluxon into the storage loop of the RSFFs, thereby increasing $I_{control}$ and producing an output voltage $V_{out}$ of roughly 5 mV. 
Conversely, a $reset$ signal removes the stored flux and turns off $I_{control}$. 
A non-return-to-zero (NRZ) signaling scheme can be implemented by driving the $set$ and $reset$ inputs with a toggle flip-flop (TFF) \cite{hashimoto2007implementation}. 

In addition to RSFF, SQUID stacks can also be driven using a Josephson transmission line (JTL) \cite{herr2010high,soloviev2007high}, an SFQ-to-DC converter \cite{khabipov2006development,razmkhah2021compact}, or a combination of both approaches \cite{mukhanov1997josephson,yoshikawa2001component,inamdar2009superconducting,tarutani2001interface}. 
SQUID stack circuits implemented using the RSFQ logic style \cite{likharev1991rsfq} typically dissipate on the order of 100-200 \textmu W \cite{gupta2007digital,hirayama2002characteristics,zhao2025high_magnetometer}, which is comparable to the power dissipation of Suzuki stacks \cite{van201364,ortlepp2013design,ortlepp2013high,konno2017fully,hironaka2025josephson,aghighi2018level,wei2010new,mustafa2023suzuki,mustafa2024dc}. 
Its power dissipation can be further reduced by employing energy-efficient logic families such as ERSFQ \cite{kirichenko2011zero}, eSFQ \cite{mukhanov2011energy}, and RQL \cite{herr2011ultra}. 
For example, an RQL-based 32-stage SQUID stack has been demonstrated in \cite{egan2022true}, exhibiting a power dissipation of 320 nW.

In a symmetric DC SQUID cell, both JJs are shunted with identical resistors and share the same $\beta_c$ parameter. 
An asymmetrical DC SQUID, where one of the JJs is unshunted and the other one is overdamped, can achieve a larger output voltage response as compared to a symmetrical design. 
SQUID stacks with asymmetrical DC SQUIDs have been presented in \cite{herr2007inductive,herr2010high,zhao2025high,zhao2025high_magnetometer}. 
A double-SQUID design has been proposed in \cite{morooka1997design}, where two DC SQUIDs share a common sensing inductor, achieving a two-fold increase in the output voltage. 
SQUID stacks with double-SQUIDs have been proposed in \cite{higuchi2019design,mizugaki2019enhanced}. 
A 4-JJ SQUID circuit housing two independent input transformers has been presented in \cite{egan2022true}, leveraging a unique switching mechanism inherent to AC-biased RQL logic. 
A SQIF (superconducting quantum interference filter)-based driver has been presented in \cite{kornev2007development} where each DC SQUID stage has a different loop size for broad band applications.

One of the challenges in designing SQUID stack circuits is that increasing the number of DC SQUIDs connected in series does not lead to a linear increase in the output voltage.
To overcome such saturation and further increase the output voltage, SQUID stacks circuit can be connected in parallel, as suggested in \cite{yoshikawa2001component}. 
Nevertheless, such a design requires additional control circuitry and results in a significantly larger layout area. 
To ensure the highest output voltage, the SQUID stacks require a careful (symmetric) layout design.
Ideally, the control signals for each DC SQUID in the array should be synchronized and applied simultaneously to achieve constructive interference. 
To achieve this, an H-tree structure was proposed in \cite{inamdar2009superconducting} to equalize the propagation delay of the control signals.

State-of-the-art SQUID stack output drivers are capable of producing an output voltages of at least 5 mV at data rates of 20-30 Gbps~\cite{zhao2025high,zhao2025high_magnetometer,gupta2019digital,mustafa2025spam}. 
With these capabilities, SQUID stacks are commonly employed in digital data links operating from 4 K to higher temperature stages  to drive both semiconductor amplifiers \cite{gupta2019digital} and photonic circuits (\textit{e.g.}, electro-optical modulators) \cite{shen2024photonic,yin2021electronic,yin2026fully}. 
Additionally, SQUID stack structures are commonly used in superconducting digital-to-analog (D/A) converters as a voltage multiplier circuit \cite{castellanos2021single,hirayama2002characteristics,hirayama2003characteristics}.

\textbf{SFQ-to-DC converter}: An SFQ-to-DC converter was presented in 1989 by Kaplunenko \textit{et al.} \cite{kaplunenko1989experimental}. 
The schematic of an SFQ-to-DC converter circuit is depicted in Fig. \ref{fig:JJ_based_drivers_schematics}(c). 
The circuit consists of a single DC SQUID directly connected to the quantizing inductance of a TFF \cite{kaplunenko1989experimental,likharev1991rsfq}.
Unlike in SQUID stack, the DC SQUID in this circuit has no magnetic coupling. 
The switching of the TFF induces a $\pi$-phase shift across the two JJs with $I_{c4}$ and $I_{c5}$ (Fig. \ref{fig:JJ_based_drivers_schematics}(c)) effectively modulating the critical current of the SQUID \cite{kaplunenko1989experimental}. 
The operating principle of an SFQ-to-DC converter is illustrated in Fig. \ref{fig:JJ_based_drivers_waveforms}(b). 
Each input SFQ pulse toggles the output voltage $V_{out}$ between high and low states, implementing an NRZ signaling scheme. 

Among JJ-based output drivers, an SFQ-to-DC converter provides the lowest power dissipation and highest data rates.
However, it produces the lowest output voltage. 
For example, the SFQ-to-DC converter presented in \cite{gupta2019digital} can produce a 1.3 mV signal at 40 Gbps while dissipating 13.4 \textmu W. 
Additionally, its power dissipation can be reduced down to 1.3 \textmu W using ERSFQ logic \cite{gupta2019digital}. 
This output driver is included in the standard cell library of various SFQ logic families \cite{sunysb_rsfq_cell_library,supertools_rsfq_cell_library,tanaka2025low,maezawa2004design,cong2024superconductor} and is typically available in the process design kit (PDK). 
Similar to SQUID stacks, SFQ-to-DC converters are commonly used in digital data links operating from 4 K to higher temperature stages \cite{gupta2019digital,mukhanov2013development,chen2025signal}. 
Due to the lower output voltage as compared to SQUID stack, an SFQ-to-DC converter requires a more complex setup for semiconductor amplifiers \cite{gupta2019digital} or a more sensitive photonic circuit (\textit{e.g.}, VCSEL) \cite{mukhanov2013development}. 
Additionally, SFQ-to-DC converter structures can also be utilized as precise current generators to manipulate and bias superconducting qubits \cite{jokar2022digiq}.

\textbf{Voltage doubler}: A voltage doubler (a.k.a. quasi-digital voltage amplifier (QDVA)) was presented in 1991 by Kaplunenko \textit{et al.} in \cite{kaplunenko1991single}. 
This voltage doubler consists of two DC SQUIDs that are strongly coupled via an inductance $L_3$ as depicted in Fig. \ref{fig:JJ_based_drivers_schematics}(e). 
When an SFQ pulse is applied at the input terminal $V_{in}$, a flux quantum and anti-flux quantum pass through the bottom and top SQUIDs producing two SFQ pulses, which appear at the output terminal $V_{out}$ \cite{kaplunenko1991single}. 
The operating principle of the voltage doubler is illustrated in Fig. \ref{fig:JJ_based_drivers_waveforms}(c). 
A train of SFQ pulses with 100 GHz frequency is periodically applied. 
The average $V_{out}$ generated is approximately two times larger than the average $V_{in}$. 

The voltage doubler output driver can be used with an SFQ-to-DC converter as suggested in \cite{ortlepp2009superconductor,wuensch2009cryogenic}. 
In this configuration, the output voltage of an SFQ-to-DC converter is doubled with a voltage doubler (\textit{e.g.}, from 200 \textmu V from SFQ-to-DC converter is increased to 438 \textmu V with the voltage doubler \cite{ortlepp2009superconductor}). 
By repeating the voltage doubler structure in vertical and horizontal directions, a larger factor of voltage multiplication can be achieved as proposed in \cite{golomidov1992single}.
In \cite{golomidov1992single}, a six-stage parallel QDVA is demonstrated to achieve a five-fold multiplication of the input signal amplitude. 

A similar concept of doubling the output voltage was introduced in \cite{herr2005stacked}, where an SFQ pulse is converted into a double flux quantum (DFQ) pulse.
By stacking several DFQ drivers in series, an output voltage of 1.30 mV can be produced at an estimated data rate of 10 Gbps \cite{herr2005stacked}. 
To generate an SFQ pulse train, a voltage-controlled oscillator (VCO) and a non-destructive readout (NDRO) circuit are employed in \cite{herr2005stacked}. 
In addition to the digital data transmission, stacked DFQ-based drivers are used in D/A converters for metrological applications.
For example, a 20-fold DFQ amplifier has achieved a maximum output voltage of 2.90 mV \cite{somei2021enhanced}.
A 1000-fold DFQ amplifier that achieves an output voltage of 43 mV has been presented in \cite{mizugaki20191000}.

\subsection{Multi-terminal devices}\label{subsection:multi_terminal_drivers}

Superconducting multi-terminal devices have three or more terminals that resemble a semiconductor transistor.
Multi-terminal devices suitable for use as output drivers can be classified into two main categories: cryotrons and superconductor-based transistors (Fig. \ref{fig:Cryostat_block_diagram_and_taxonomy}(b)). 

\subsubsection{\textbf{Cryotron}}
Cryotrons \cite{buck2007cryotron} consist of a superconducting channel through which current flows between the drain and source terminals, and an input terminal that receives the control signal to modulate the channel characteristics (\textit{i.e.}, switching it to the resistive state).
The following subsections describe several variations of cryotron-based output drivers.

\begin{figure*}
  \centering
    \includegraphics[width=0.8\textwidth]{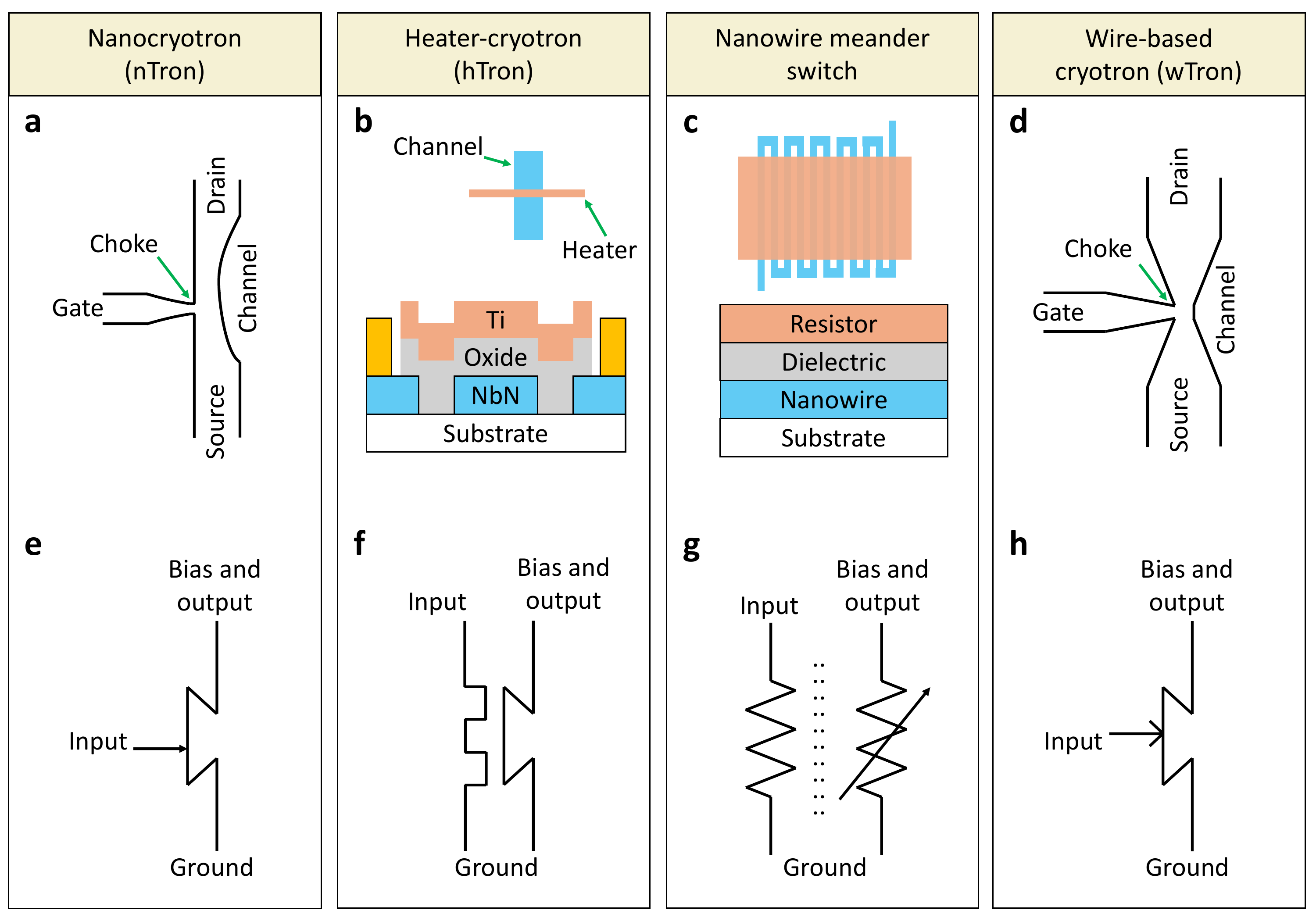}
    \caption{Multi-terminal devices. Layout view/structure of 
    \textbf{a}, nanocryotron (nTron) \cite{mccaughan2014superconducting},
    \textbf{b}, multi-layered heater-cryotron (hTron) \cite{baghdadi2020multilayered},
    \textbf{c}, nanowire meander switch \cite{mccaughan2019superconducting}, and
    \textbf{d}, wire-based cryotron (wTron) \cite{paul2025photolithography}. 
    The circuit symbols of 
    \textbf{e}, nTron,
    \textbf{f}, multi-layered hTron,
    \textbf{g}, nanowire meander switch, and
    \textbf{h}, wTron.
    }
    \label{fig:Multi_terminal_devices}
\end{figure*}

\textbf{Nanocryotron (nTron)}: 
nTron was proposed in 2014 by McCaughan and Berggren \cite{mccaughan2014superconducting}. 
The layout view and circuit symbol of nTron are depicted in Fig. \ref{fig:Multi_terminal_devices}(a) and Fig. \ref{fig:Multi_terminal_devices}(e), respectively. 
An nTron is implemented with a single superconductor layer (\textit{e.g.}, NbN \cite{mccaughan2014superconducting,zhao2017nanocryotron} or NbTiN \cite{tanaka2017josephson,sano2018thermally}) that is shaped in a specific way, as shown in Fig. \ref{fig:Multi_terminal_devices}(a). 
Particularly, the nTron has an input gate that injects a small amount of current, creating a localized, Joule-heat hotspot in the choke area (Fig. \ref{fig:Multi_terminal_devices}(a)). 
As a result of the rapid transition from the superconducting to the resistive state, an nTron can drive large impedance loads (more than 100 k$\Omega$) \cite{mccaughan2014superconducting}.

An nTron can be triggered with SNSPD pulses \cite{mccaughan2014superconducting}. 
Additionally, it can interface directly with SFQ circuits, as experimentally demonstrated in \cite{zhao2017nanocryotron}, enabling seamless integration without the need for intermediate signal conditioning stages. 
As compared to JJ-based output drivers, an nTron offers the largest output voltage ranging from 100 mV up to 8.1 V \cite{sano2018thermally,tanaka2017josephson,mccaughan2014superconducting}), while simultaneously achieving quite low power dissipation (around 8 nW \cite{zhao2017nanocryotron}), and extremely compact layout footprint of roughly 0.1 \textmu m$^2$, excluding the bias inductors \cite{zhao2017nanocryotron}. 
However, the primary limitation is the low switching frequency of around 1 GHz, which arises from the relatively long transition times required for the device to switch between superconducting and resistive states \cite{zhao2017nanocryotron}. 
Additionally, an nTron can be configured to operate in either self-reset or latched modes and can therefore be biased using a DC or AC power supply, respectively \cite{zhao2017nanocryotron}.

The nTron fabrication process is based on SNSPD technology \cite{paul2025photolithography}. 
Therefore, nTrons can be integrated on the same chip with SNSPDs and employed for their readout \cite{zheng2019characterize,castellani2024nanocryotron}. 
In the case of interfacing SFQ circuits, nTrons typically need to be fabricated on a separate chip and then, for example, wire bonded to the SFQ and CMOS chips \cite{sano2018thermally,tanaka2017josephson,zhao2017nanocryotron}. 
An nTron can also serve as an output driver for 4 K Josephson-CMOS hybrid memory, as suggested in \cite{tanaka2017josephson}. 
It is estimated that a Josephson-CMOS hybrid memory incorporating nTrons could achieve roughly 1/12 of the power dissipation of conventional Suzuki stack-based hybrid memory architectures \cite{tanaka2017josephson}.
Nevertheless, nTron fabrication techniques require further development along with more advanced circuit-level simulation models \cite{sano2018thermally,yasukawa2024ginzburg}.

\textbf{Heater-cryotron (hTron)}: An hTron was presented in 2018 by Zhao \textit{et al.} \cite{zhao2018compact}.
The first version of the hTron had a planar structure with two isolated and closely spaced (40 nm spacing) nanowires \cite{zhao2018compact}. 
A multi-layered hTron was presented later in \cite{baghdadi2020multilayered}. 
The physical layout and circuit symbol of a multi-layered hTron are depicted in Fig. \ref{fig:Multi_terminal_devices}(b) and Fig. \ref{fig:Multi_terminal_devices}(f), respectively. 
The device consists of channel and heater nanowires, making it a four-terminal device. 
Due to the electrical isolation of the heater (\textit{i.e.}, input) and the channel, an hTron does not experience a current leakage problem, unlike an nTron \cite{baghdadi2020multilayered}. 
When an input signal is applied to the heater, it transitions from the superconducting state to the resistive state. 
Due to the local temperature increase caused by Joule heating and the close proximity of the heater, the channel's critical current is suppressed, causing it to switch to the resistive state. 

hTrons have been employed in the design of superconducting memory elements \cite{zhao2018compact,nguyen2020cryogenic}, cryogenic logic cells \cite{alam2023cryogenic}, and SNSPD readout circuits \cite{baghdadi2020multilayered}. 
Similar to nTrons, hTrons are capable of producing large output voltages on the order of 500-600 mV \cite{nguyen2020cryogenic,baghdadi2020multilayered}. 
A SPICE model for the hTron device is proposed in \cite{karam2025parameter}.

\textbf{Nanowire meander switch}: A nanowire meander switch has been proposed in \cite{mccaughan2019superconducting}. 
This device has a similar structure and switching mechanism as an hTron, as illustrated in Fig. \ref{fig:Multi_terminal_devices}(c) with the corresponding circuit symbol depicted in Fig. \ref{fig:Multi_terminal_devices}(g). 
In this four-terminal device, the channel is a nanowire meander and the heater is a wide resistor line. 
The nanowire meander switch can deliver a 1.12 V output voltage, which is sufficient to drive a cryogenic light-emitting diode (LED), as demonstrated in \cite{mccaughan2019superconducting}. 
The switch exhibits a turn-on time of less than 300 ps and a turn-off time of approximately 15 ns.
This relatively long thermal recovery time of the switch can be advantageous in optoelectronic applications (\textit{e.g.}, driving LEDs and modulators) and neuromorphic circuits \cite{shainline2018circuit}.

\textbf{Wire-based cryotron (wTron)}: 
A wTron was presented in 2025 by Paul \textit{et al.} \cite{paul2025photolithography}. 
This device has a similar structure and switching mechanism as nTron, as illustrated in Fig. \ref{fig:Multi_terminal_devices}(d). 
The corresponding circuit symbol of wTron is depicted in Fig. \ref{fig:Multi_terminal_devices}(h). 
While an nTron can be fabricated at the sub-100 nm scale, enabling monolithic integration with SNSPDs, a wTron is typically fabricated at the micrometer scale, making it compatible with SFQ-based fabrication processes. 
For example, wTron devices have been fabricated in MIT Lincoln Lab SFQ5ee process in \cite{paul2025photolithography}. 
The larger dimensions of a wTron enable it to drive high capacitive loads (up to 500 pF) typical of semiconductor (CMOS) circuits \cite{paul2025photolithography}.

\subsubsection{\textbf{Superconductor-based transistor}} 
While cryotrons function primarily as digital switches, several superconductor-based devices exhibit analog behavior analogous to that of semiconductor transistors. 
Examples include the quiteron \cite{faris2003quiteron,frank1984circuit}, superconducting field-effect transistor (FET) \cite{nishino1989sufet}, superconducting-ferromagnetic transistor \cite{nevirkovets2009superconducting,nevirkovets2014superconducting,nevirkovets2020characterization,nevirkovets2023electrically}, Josephson FET \cite{akazaki1996josephson,de2019josephson,wen2019josephson}, superconducting Dayem bridge FET \cite{paolucci2018ultra}, metallic supercurrent FET \cite{de2018metallic}, and superconducting quantum interference proximity transistor \cite{giazotto2010superconducting}. 
Although there are currently no demonstrations of SFQ-based or SNSPD-based triggering of superconductor-based transistors, these devices have the potential to be integrated into more complex circuits and serve as second stage amplifiers, such as a low noise amplifier or a differential amplifier, similar to those used in semiconductor circuits \cite{van201364,konno2017fully,gupta2013low,ravindran2014power,ravindran2017energy,chen2025sfq}). 
Nevertheless, to the best of the authors' knowledge, no digital data link has yet been implemented using superconductor-based transistors.

\section{Effect of Fabrication Technology}\label{section:fabrication}

In this section, the effects of different fabrication technology parameters on the performance of interface circuits are discussed. 
Particular attention is given to JJ-based output drivers and semiconductor amplifiers due to the widespread availability of various fabrication processes for these interface circuits.

\subsection{JJ-based output drivers}

JJ-based fabrication processes (such as MIT Lincoln Lab SFQ5ee \cite{tolpygo2016advanced}, SEEQC SFQ-C5SL \cite{yohannes2023high}, ADP2 \cite{nagasawa2014nb}, SIMIT Nb03P \cite{ying2021development}, and FLUXONICS CJ2 \cite{kunert2024advanced}) are characterized primarily by the critical current density $J_c$.
Typical values of $J_c$ range from 10 to 100~\textmu A/\textmu m$^2$. 
The switching speed of JJs scales approximately with the square root of the critical current density, $J_c$, \cite{inamdar2009superconducting}. 
For example, increasing the critical current density $J_c$ from 10~\textmu A/\textmu m$^2$ to 100~\textmu A/\textmu m$^2$ results in approximately a 3.16-fold increase in switching speed.
Therefore, processes with higher $J_c$ values (\textit{e.g.}, 100~\textmu A/\textmu m$^2$) are preferred for high-speed applications of JJ-based drivers, such as SQUID stacks and SFQ-to-DC converters. 

In addition to the switching speed, the output voltage of overdamped JJs also scales approximately with the square root of $J_c$ \cite{inamdar2009superconducting}. 
SQUID-based output drivers should be fabricated using a high $J_c$ process to maximize the output voltage swing. 
The output voltage of latching drivers (with underdamped JJs) depends on the gap voltage $V_g$, which is relatively consistent across modern fabrication processes and typically ranges from 2.6 to 2.8 mV. 

Low-power applications such as superconducting qubit control and readout often use processes with lower $J_c$ \cite{yohannes2023high,tanaka2025low}.
For instance, an SFQ-to-DC converter fabricated using the AIST 2.5~\textmu A/\textmu m$^2$ process dissipates 74 nW of power, which is approximately two orders of magnitude lower than that of a comparable SFQ-to-DC converter designed with MIT Lincoln Lab 100~\textmu A/\textmu m$^2$ process \cite{gupta2019digital}. 

The majority of JJ-based output drivers incorporate explicit inductors (\textit{i.e.}, not parasitic inductors) in their circuit design, as illustrated in Fig. \ref{fig:JJ_based_drivers_schematics}. 
Additionally, low-power biasing schemes such as ERSFQ and eSFQ, applicable to SQUID-based output drivers, employ large bias inductors on the order of hundreds of picohenries \cite{kirichenko2011zero,mukhanov2011energy}.
These inductors require a significant layout footprint, thereby limiting the scalability of output drivers.
Inductor-less superconductor electronic circuits have been proposed in the literature, utilizing conventional and bistable JJs (\textit{e.g.}, 2-$\phi$ JJ) \cite{soloviev2021superconducting,jabbari2023all,cong2024superconductor,elmitwalli2023bistable,salameh2022superconductive,mitrovic2025josephson}. 
Similar approaches could be applied to SQUID-based output drivers to reduce or potentially eliminate the need for inductors. 
However, since bistable JJs incorporate ferromagnetic materials, the corresponding fabrication processes remain immature for large-scale circuit integration \cite{mitrovic2025josephson}. 

Latching output drivers are sensitive to the parasitic capacitance to ground ($C_g$) associated with each JJ. 
A lower $C_g$ is preferred in these drivers to minimize delay and reset time \cite{liu2007josephson}. 
In Suzuki stack interface circuits, the ground plane underneath the drivers is removed to reduce $C_g$, as suggested in \cite{ortlepp2013design}. 

JJ-based output drivers are typically designed to operate at 4 K or below due to the Nb-based fabrication process, which has a critical temperature around 9 K.
A high-temperature superconductor (HTS)-based fabrication process, which uses YBCO and LSAT materials, has been used to fabricate latching output drivers in \cite{hato2003high,hato2003sfq,hato2005output}. 
To achieve an underdamped behavior, the JJs are shunted with capacitors, and a $J_c$ of 200~\textmu A/\textmu m$^2$ is used to reduce the punchthrough probability of JJs \cite{hato2003high}. 
In \cite{hato2003high}, a 4-JJ Suzuki stack circuit produces 1.0 mV at 30 K. 
The measured output voltage is smaller than expected (\textit{i.e.}, only one of the JJs in the stack is turned on), which could be caused by the greater than 30\% spread of the critical current of JJs \cite{hato2003high}.
Further optimization of the HTS-based process is required to reduce the spread of critical current \cite{hato2003high}. 
Superconducting output drivers operating at intermediate temperature stages (\textit{e.g.}, 10-50 K) could significantly improve the cooling power budget of the cryostat as compared to semiconductor amplifiers.

\subsection{Semiconductor amplifiers}

Semiconductor amplifiers in cryogenic digital data links are typically implemented using CMOS or bipolar CMOS (BiCMOS) technologies.
CMOS-based amplifiers have been used in 4 K Josephson-CMOS hybrid memories, as reported in \cite{van201364,konno2017fully}.
High-frequency data links from 4 K to room-temperature electronics utilize BiCMOS-based amplifiers, which can provide faster switching speed (\textit{i.e.}, data rate), and  higher gain and output drive current as compared to CMOS technology. 
Examples of SiGe BiCMOS amplifiers have been demonstrated in \cite{gupta2013low,ravindran2014power,ravindran2017energy,chen2025sfq}. 

Existing demonstrations of cryogenic semiconductor amplifiers within digital data links often utilize mature fabrication nodes such as 180 nm CMOS \cite{jin2012investigation,kuwabara2013design,konno2017fully}, 65 nm CMOS \cite{wei2010new}, and 130 nm BiCMOS \cite{gupta2013low,ravindran2014power,ravindran2017energy,chen2025sfq} primarily due to relatively low fabrication costs. 
Konno \textit{et al.} \cite{konno2017fully} discuss the possibility of using modern fabrication nodes ranging from 22 nm to 7 nm FinFET technology to reduce power dissipation and improve switching speed in Josephson-CMOS memory systems, which include semiconductor amplifiers. 
Additionally, the threshold voltage of MOSFETs reduces in more advanced process nodes \cite{salman2012high}. 
By  lowering the threshold voltage of the second stage semiconductor amplifier, the required output voltage of the first stage superconducting output driver can be reduced, leading to lower overall power dissipation. 

A hybrid amplifier that uses both CMOS and JJ devices was proposed in 1993 by Ghoshal \textit{et al.} \cite{ghoshal1993superconductor}. 
This hybrid amplifier employs an inverter circuit comprising an NMOS transistor and a series stack of JJs, which together replace the PMOS transistor \cite{ghoshal1993superconductor}. 
The JJ stack provides a negative differential resistance and reduces the required time to charge the load capacitor \cite{ghoshal1993superconductor,ghoshal1995cmos}. 
Additionally, this JJ stack eliminates the Miller effect \cite{ghoshal1993superconductor,ghoshal1995cmos}.
Similar hybrid amplifiers have been presented in \cite{feng2003josephson,liu2007latency}. 
An experimental demonstration of a hybrid amplifier (with three NMOS transistors and 400-JJ stack) has been presented in \cite{fujiwara2010new}, where CMOS and Josephson chips have been fabricated separately and interconnected via solder-bump bonding.

\section{Comparison of Superconducting Output Drivers}\label{section:comparison}

In this section, the superconducting output drivers are compared based on circuit- and system-level parameters. 
Particularly, the parameters such as output voltage, data rate, power dissipation, layout area, heat (thermal) load of cryogenic cables, line coding scheme, flux trapping, bit-error rate, and hardware security are considered.

\subsection{Circuit-level parameters}

\begin{figure*}
  \centering
    \includegraphics[width=1\textwidth]{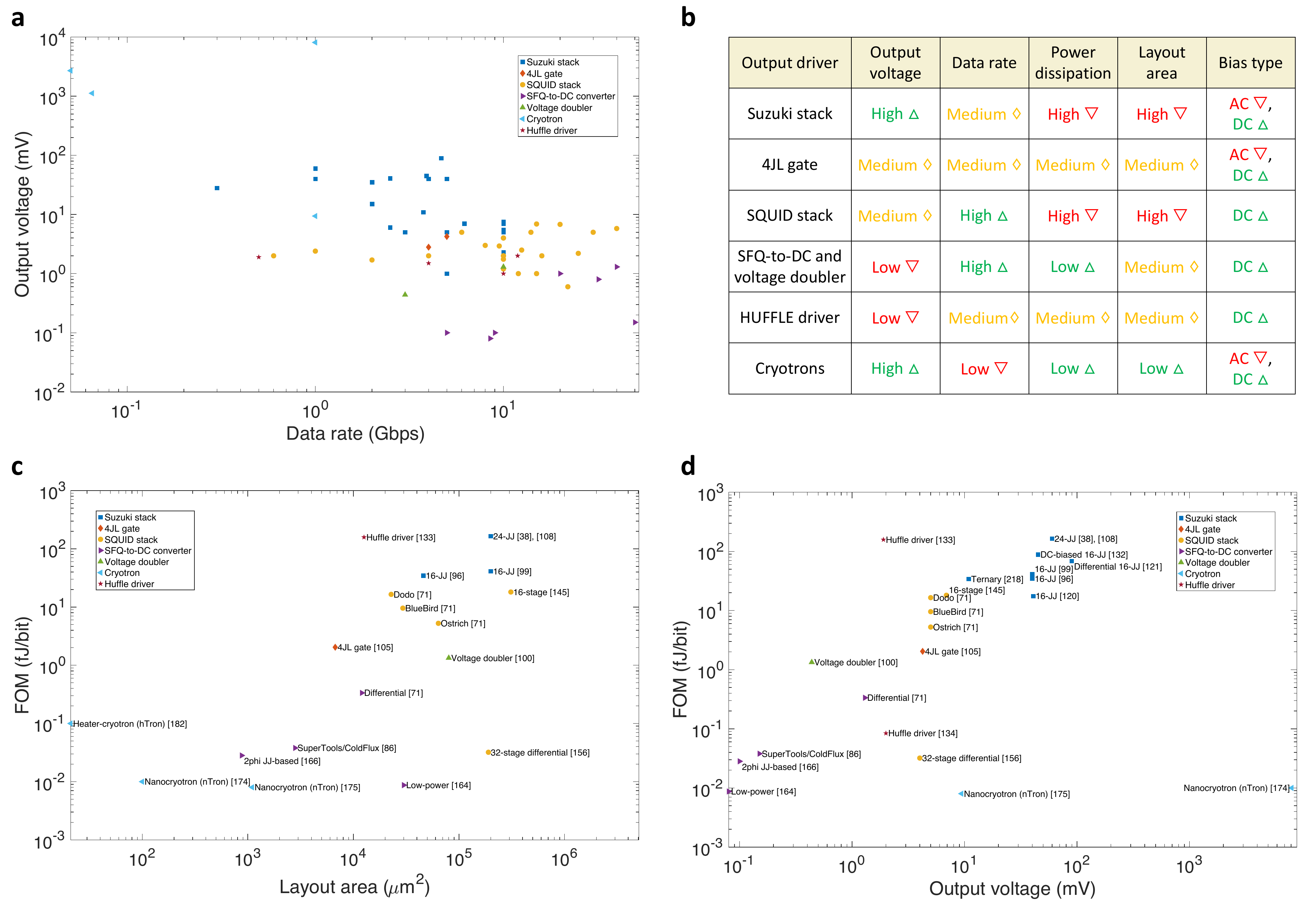}
    \caption{Comparison of superconducting output drivers. \textbf{a}, Output voltage and data rate parameters comparison. 
    \textbf{b}, General trends in output voltage, data rate, power dissipation, layout area, and bias type. Green, orange, and red colors ($\bigtriangleup$, $\Diamond$, and $\bigtriangledown$ symbols) correspond to desired, intermediate, and undesired performance, respectively.
    \textbf{c}, FOM and layout area parameters comparison.
    \textbf{d}, FOM and output voltage parameters comparison. FOM is calculated as power dissipation (only circuit-level component) divided by data rate.
    }
    \label{fig:Comparison_plots}
\end{figure*}

Due to the stringent cooling power constraints at cryogenic temperatures, the superconducting output drivers with higher data rate and greater output voltage are highly desirable.
A higher data rate per channel reduces the number of cryogenic interconnects required. 
A higher output voltage from the first stage output driver enables the second stage semiconductor amplifier to operate with lower gain and reduced power dissipation. 
A comparison of superconducting output drivers in terms of output voltage and data rate parameters, as reported in the literature over the past 40–50 years, is shown in Fig. \ref{fig:Comparison_plots}(a).
As summarized in the figure, no output driver has demonstrated both high output voltage and high data rate concurrently.
Therefore, there is a trade-off between the output voltage and the data rate. 
For example, Suzuki stacks deliver output voltages on the order of tens to hundreds of mV and operate at data rates of a few Gbps. 
In contrast, SQUID stacks achieve data rates of 20-30 Gbps with output voltages below 10 mV. 
Cryotron-based devices deliver the highest output voltages, exceeding 1 V, but operate at data rates below 1 Gbps. 
SFQ-to-DC converters produce the lowest output voltages, around 1 mV, while supporting the highest data rates of up to 40-50 Gbps.

In addition to output voltage and data rate, power dissipation and layout area are critical factors in evaluating output drivers. 
Lower power dissipation is desirable to maximize the available cooling power and support a higher number of interface channels. 
Layout area is particularly important for SFQ-based chips, where current fabrication processes impose limitations on integration density. 
Fig. \ref{fig:Comparison_plots}(b) summarizes the key trends in these parameters.
The power dissipation of Suzuki and SQUID stack drivers is the highest and exceeds 100 \textmu W per channel. 
Although Suzuki stack has a significantly lower number of JJs as compared to SQUID stack, this driver has a large bias resistor (see $R_b$ in Fig. \ref{fig:JJ_based_drivers_schematics}(d)), which is needed to maintain the high output voltage during switching \cite{ortlepp2013design}. 
$R_b$ contributes to both power dissipation and layout area \cite{ortlepp2013design,mustafa2022optimization} and could possibly be reduced with a DC-biased scheme and by adding a large bias inductor as suggested in \cite{mustafa2024dc}. 
The power dissipation of a SQUID stack (and other SQUID-based output driver) depends on the type of SFQ logic being used. 
For example, RSFQ logic family has large static power dissipation due to the resistive bias scheme \cite{likharev1991rsfq}. 
Energy-efficient variants such as the RQL-based SQUID stack has lower power dissipation on the order of sub-\textmu W, as demonstrated in \cite{egan2022true}. 
Fig. \ref{fig:Comparison_plots}(c) and Fig. \ref{fig:Comparison_plots}(d) present comparisons of superconducting output drivers based on a figure of merit (FOM), layout area, and output voltage. 
This FOM, proposed by Gupta \textit{et al.} \cite{gupta2019digital}, is defined as the ratio of power dissipation to data rate.
Each category of output drivers (see taxonomy in Fig. \ref{fig:Cryostat_block_diagram_and_taxonomy}(b)) is represented as a distinct group in Fig. \ref{fig:Comparison_plots}(c) and Fig. \ref{fig:Comparison_plots}(d).
From Fig. \ref{fig:Comparison_plots}(d), the trade-off between output voltage and the figure of merit (FOM) is clearly observed for JJ-based output drivers.
Specifically, higher output voltages correspond to higher FOM values, that is, increased power dissipation and/or reduced data rates.

\subsection{Heat load of cryogenic cables}
Cryogenic bias and signal cables contribute a significant amount of heat (thermal) load, 
which reduces the cooling power available for the cryostat payload \cite{krinner2019engineering,zhuldassov2025cryogenic}. 
Depending on the type of power supply (\textit{i.e.} AC or DC), different cryogenic cables are preferred.
Coaxial cables are typically used for AC bias, whereas twisted pair cables are used for DC bias power supply \cite{mustafa2024dc}. 
The heat load of AC-bias cables is generally higher as compared to the DC-bias cables \cite{klostermann1991heat}. 
For instance, Mustafa \textit{et al.} \cite{mustafa2024dc} compared the heat load of the bias cables for AC- and DC-biased 4JL gates and Suzuki stack circuits. 
AC-bias cables have approximately three orders of magnitude higher heat load as compared to the DC-biased cables (\textit{e.g.}, for a 30-channel Suzuki stack interface, an AC-bias cable has 66.0 \textmu W, and a DC-bias cable has 34.6 nW) \cite{mustafa2024dc}. 
The type of biasing (AC and/or DC) for different superconducting output drivers is listed in Fig. \ref{fig:Comparison_plots}(b). 

The cryogenic signal cables are used to transmit the digital data from the superconducting output drivers. 
Several BeCu-Cu-BeCu stripline ribbon cables, which are employed for 4 K to room-temperature data transmission, have been analyzed and compared in \cite{gupta2019digital}. 
These cables have a 4 K heat load of 440 \textmu W per line at 2.1 Gbps, 740 \textmu W per line at 5.7 Gbps, and 1118 \textmu W per line at 10.0 Gbps. 
Additionally, BeCu coaxial cables have been reported with 2473 \textmu W per line at 42.9 Gbps \cite{gupta2019digital}. 
Superconducting output drivers have two types of output signaling: single-ended and differential. 
Differential signaling offers a larger output voltage swing and improved noise immunity due to the common-mode rejection property \cite{gupta2019digital}. 
Superconducting output drivers with differential signaling have been proposed in the literature including implementations based on SQUID stacks \cite{inamdar2009superconducting,gupta2019digital}, SFQ-to-DC converters \cite{inamdar2009superconducting,gupta2019digital}, and Suzuki stacks \cite{mustafa2023suzuki}. 
The single-ended and differential output drivers require one and two output cables, respectively.
As a result, the heat load of a differential output driver is approximately twice that of a single-ended driver \cite{gupta2019digital}.

\subsection{Line coding schemes}

Superconducting output drivers typically employ binary data transmission. 
A binary bit string is encoded with either NRZ-M (non-return-to-zero mark) or RZ (return-to-zero) line coding schemes. 
In the NRZ-M scheme, the output signal is toggled from high-to-low or low-to-high voltage state when a logical `1' is transmitted. 
A logical `0' corresponds to no transition, where the previous voltage state is maintained. 
The NRZ-M scheme is commonly used in SQUID-based output drivers such as the SQUID stack, SFQ-to-DC converter, and voltage doubler, as shown in Fig. \ref{fig:JJ_based_drivers_waveforms}(a)-(c). 
Additionally, the HUFFLE driver can produce an NRZ-M signaling (Fig. \ref{fig:JJ_based_drivers_waveforms}(d)). 
Other types of output drivers (\textit{e.g.}, latching and cryotrons) typically employ the RZ scheme, where the logical `1' produces a high voltage state for the first half of the clock period and returns to zero voltage state in the second half. 
A logical `0' is represented as the low output voltage for the entire clock period. 
The RZ line coding of Suzuki stack and 4JL gate circuits can be observed in Fig. \ref{fig:JJ_based_drivers_waveforms}(e) and Fig. \ref{fig:JJ_based_drivers_waveforms}(f), respectively. 

SQUID-based output drivers can also be be operated in the RZ line coding scheme by utilizing a delay element as demonstrated in \cite{hashimoto2007implementation}. 
However, the NRZ scheme is generally preferred in high-speed digital data links because it provides better signal quality at higher data rates as compared to the RZ scheme \cite{koch1999nrz,herr2005stacked}. 

Multi-level signal transmission increases the effective data rate by encoding more than one bit per clock cycle. 
A ternary superconducting output driver has been proposed in \cite{mustafa2025ternary}. 
This ternary driver builds upon a re-designed Suzuki stack circuit with three arrays of serially connected JJs and can produce three output voltage levels \cite{mustafa2025ternary}. 
For a fixed clock frequency, the ternary digital data link can increase the effective data rate by 50\% as compared to a binary data link \cite{mustafa2025ternary}. 
Alternatively, a ternary link can potentially reduce the number of cryogenic cables by 33\% while maintaining the same effective data rate \cite{mustafa2025ternary}. 
A similar concept of a multi-level output driver has been proposed in \cite{mustafa2025spam}, where the SQUID stack circuit is configured to produce a 4-level pulse-amplitude modulated (PAM-4) signal.
With four output voltage levels, the effective data rate can be doubled \cite{mustafa2025spam}. 
The PAM-4 signals generated by the SQUID stack driver can be directly interfaced with modern cryogenic FPGAs (\textit{e.g.}, Xilinx Artix-7), memory (\textit{e.g.}, Micron's GDDR6X), and serial links (\textit{e.g.}, Intel's CEI-56G), all of which support PAM-based communication protocols \cite{mustafa2025spam}.

\subsection{Flux trapping}

Superconductor digital circuits are susceptible to flux trapping caused by external magnetic fields, which occurs when magnetic flux becomes pinned in defects or inhomogeneities within the superconducting material.
This trapped flux can lead to localized disruptions in superconductivity, resulting in performance degradation or circuit failure \cite{robertazzi1997flux,fourie2021experimental}.
This effect may occur in SQUID-based, HUFFLE, and 4JL gate output drivers due to the superconducting loops in their implementations (Fig. \ref{fig:JJ_based_drivers_schematics}). 
As previously mentioned, a Suzuki stack driver incorporates two small resistors $R_{br}$ (see Fig. \ref{fig:JJ_based_drivers_schematics}(d)), which prevent flux trapping \cite{ortlepp2013design}. 
Cryotron devices do not have superconducting loops and, therefore, are not susceptible to flux trapping.

\subsection{Bit-error rate (BER)}

Bit errors within digital data links can arise from fabrication defects, process parameter variations, flux trapping, signal attenuation, and noise. 
The bit-error rate (BER) is determined by dividing the number of bit-errors  by the total number of transmitted bits. 
State-of-the-art output drivers based on SQUID stacks (data rate of 10-20 Gbps) \cite{gupta2019digital,hashimoto2007implementation,hashimoto2009measurement}, SFQ-to-DC converters (data rate of 18 Gbps) \cite{gupta2019digital}, and Suzuki stacks (data rate of 1-5 Gbps) \cite{ortlepp2013design,ortlepp2013high,harada2004josephson,kojima2005parameter} have demonstrated BER below 10$^{-12}$.
The early implementation of nTron-based output driver has the BER of 4$\times$10$^{-4}$ operating at around 100 Mbps \cite{zhao2017nanocryotron}.

Error-correction codes can be utilized to detect and correct bit errors. 
Linear block codes have been proposed for superconducting output drivers in \cite{peng2019solution,mustafa2025lightweight}. 
By employing, \textit{e.g.}, a (32,32) linear block code, BER can be reduced by at least three orders of magnitude \cite{peng2019solution}.

\subsection{Hardware security}

Hardware security of integrated circuits is a research field dedicated to uncovering hardware-level vulnerabilities and preventing malicious activities such as intellectual property piracy, reverse engineering, covert communication, and side-channel attacks. 
Unlike software-level attacks, hardware security vulnerabilities are more difficult to mitigate once the chip has been fabricated and released to market. 
Therefore, hardware security should be considered during the design of integrated circuits along with power, performance, area, and cost metrics. 

One of the most prominent threats in hardware security comes from side-channel attacks.
Side-channel attacks monitor the tiny variations in, \textit{e.g.}, power supply, and are able to infer the internal circuit switching. 
A considerable amount of side-channel leakage has been identified in superconducting output drivers as compared to standard SFQ circuits \cite{mustafa2023side_channel,mustafa2023side_channel_SS,mustafa2024side_channelQC}. 
By measuring the power supply from room-temperature electronics, it is possible to decode the bits, which are transmitted by superconducting output drivers, without physically touching the output signal pads \cite{mustafa2023side_channel}. 
Significant fluctuations in the bias current on the order of tens to hundreds of \textmu A have been identified in SFQ-to-DC converters, SQUID stacks, and Suzuki stack output drivers \cite{mustafa2024side_channelQC}. 

Since it is possible to obtain the information from power supply variations, one may use the side-channel leakage information for cryogenic testing and verification purposes, as proposed in \cite{mustafa2024built}. 
In this approach, superconducting output drivers are connected to a common power supply, generating a measurable side-channel leakage \cite{mustafa2024built}. 
The output voltage terminals are not connected to the output pads, which significantly reduces the thermal load (\textit{i.e.}, there are no cryogenic cables connected to the output signals).

\section{Conclusion and Outlook}\label{conclusion}

In this work, a detailed review of superconductor-semiconductor interface circuits is presented.
A particular focus is placed on superconducting output drivers that can convert and amplify SFQ pulses to DC voltage signals to be processed by semiconductor circuits.
Two main categories of superconducting output drivers are identified such as JJ-based drivers and multi-terminal devices. 
The working principle, advantages, and limitations are discussed for various types of interface circuits that have been proposed in the past 40-50 years. 
A comprehensive comparison of these interface circuits demonstrates the trade-offs among several design parameters including output voltage, data rate, power dissipation, layout area, power supply type (AC or DC), and BER. 
The selection of an appropriate interface circuit requires thorough evaluation and depends on the design requirements for specific applications such as HPC, superconducting quantum computing, digital signal processing, single-photon detection, and neuromorphic computing. 
With continuous improvements in fabrication processes, it is important to consider key parameters that influence the performance of specific interface circuits. 
This survey presents such an analysis focusing on JJ-based output drivers and semiconductor amplifiers. 
Due to the relatively recent introduction of multi-terminal devices and their ongoing development, further research is necessary to understand the implications of various fabrication processes on multi-terminal superconducting output drivers.
Furthermore, the development of interface circuits that leverage novel ferromagnetic JJs, which exhibit bistable behavior, and superconductor-based transistors needs future research.

\section*{Acknowledgment}
The authors would like to thank D. Scott Holmes for fruitful discussions.

\bibliographystyle{./bibliography/IEEEtran}
\bibliography{./bibliography/references_doi}

@article{van201364,
  title={64-kb Hybrid {Josephson-CMOS} 4 {K} random-access memory with 12 {mW} read power and 400 ps access time},
  author={Van Duzer, T and Zheng, L and Whiteley, S and Kim, H and Kim, J and Meng, X and Ortlepp, T},
  journal={IEEE Transactions on Applied Superconductivity},
  volume={23},
  number={3},
  note={{A}rt. no. 1700504},
  month={June},
  year={2013}
}

@article{ortlepp2013design,
  title={Design guidelines for {Suzuki} stacks as reliable high-speed {Josephson} voltage drivers},
  author={Ortlepp, T and Zheng, L and Whiteley, SR and Van Duzer, T},
  journal={Superconductor Science and Technology},
  volume={26},
  number={3},
  note = {{A}rt. no. 035007},
  month={January},
  year={2013},
  publisher={IOP Publishing}
}

@article{bhat199910,
  title={A 10 {GHz} digital amplifier in an ultra-small-spread {high-J/sub c/Nb/Al-AlOx/Nb} integrated circuit process},
  author={Bhat, Anupama and Meng, Xiaofan and Whiteley, Stephen and Jeffery, Mark and Van Duzer, Theodore},
  journal={IEEE Transactions on Applied Superconductivity},
  volume={9},
  number={2},
  pages={3232--3235},
  month={June},
  year={1999}
}

@article{ortlepp2013high,
  title={High-speed hybrid superconductor-to-semiconductor interface circuit with ultra-low power consumption},
  author={Ortlepp, Thomas and Whiteley, Stephen R and Zheng, Lizhen and Meng, Xiaofan and Van Duzer, Theodore},
  journal={IEEE Transactions on Applied Superconductivity},
  volume={23},
  number={3},
  note = {{A}rt. no. 1400104},
  month={June},
  year={2012}
}

@article{harada2000high,
  title={A high-speed {Josephson} latching driver for a superconducting single-flux-quantum system to semiconductor system interface},
  author={Harada, Naoki and Yoshida, Akira and Yokoyama, Naoki},
  journal={Japanese Journal of Applied Physics},
  volume={39},
  number={11B},
  pages={L1158},
  month={November},
  year={2000},
  publisher={IOP Publishing}
}

@article{hato2003high,
  title={A high-temperature superconductor latching driver operated at {30 K} for a single-flux-quantum/semiconductor interface},
  author={Hato, Tsunehiro and Horibe, Masahiro and Harada, Naoki and Yoshida, Akira and Ishimaru, Yoshihiro and Tarutani, Yoshinobu and Tanabe, Keiichi and Yokoyama, Naoki},
  journal={Superconductor Science and Technology},
  volume={16},
  number={12},
  note = {{A}rt. no. 1508},
  month={November},
  year={2003},
  publisher={IOP Publishing}
}

@inproceedings{suzuki1988josephson,
  title={A {Josephson} Driver to Interface {Josephson} Junctions to Semiconductor Transistors},
  author={Suzuki, H. and Inoue, A. and Imamura, T. and Hasuo, S.},
  booktitle={{IEEE International Electron Devices Meeting}},
  pages={290--293},
  month={December},
  year={1988}
}

@article{suzuki1990applications,
  title={Applications of synchronized switching in series-parallel-connected {Josephson} junctions},
  author={Suzuki, HIDEO and Imamura, Takeshi and Hasuo, SHINYA},
  journal={IEEE Transactions on Electron Devices},
  volume={37},
  number={11},
  pages={2399--2405},
  month={November},
  year={1990}
}

@article{suzuki1999interface,
  title={An interface circuit for a {Josephson-CMOS} hybrid digital system},
  author={Suzuki, M and Maezawa, M and Takato, H and Nakagawa, H and Hirayama, F and Kiryu, S and Aoyagi, M and Sekigawa, T and Shoji, A},
  journal={IEEE Transactions on Applied Superconductivity},
  volume={9},
  number={2},
  pages={3314--3317},
  month={June},
  year={1999}
}

@article{kunert2024advanced,
  title={Advanced {FLUXONICS} Process {CJ2} Based on Sub-$\mu$m-Sized Cross-Type {Nb/AlO x/Nb Josephson} Junctions for Mixed Signal Circuits},
  author={Kunert, J and Schmelz, M and Peiselt, K and Oelsner, G and Reddy, SG and Ortlepp, T and Stolz, R},
  journal={IEEE Transactions on Applied Superconductivity},
  volume={34},
  number={3},
  note = {{A}rt. no. 1101105},
  month={May},
  year={2024}
}

@article{konno2017fully,
  title={Fully Functional Operation of Low-Power 64-Kb {Josephson-CMOS} Hybrid Memories},
  author={Konno, Gen and Yamanashi, Yuki and Yoshikawa, Nobuyuki},
  journal={IEEE Transactions on Applied Superconductivity},
  volume={27},
  number={4},
  note = {{A}rt. no. 1300607},
  month={June},
  year={2017}
}

@article{harada2002high,
  title={High-speed demonstration of an output interface driver for single-flux quantum systems},
  author={Harada, Naoki and Yoshida, Akira and Yokoyama, Naoki},
  journal={IEEE Transactions on Applied Superconductivity},
  volume={12},
  number={3},
  pages={1852--1856},
  month={September},
  year={2002}
}

@article{china2022high,
  title={A high-speed interface based on a {Josephson} latching driver for adiabatic quantum-flux-parametron logic},
  author={China, Fumihiro and Takeuchi, Naoki and Suzuki, Hideo and Yamanashi, Yuki and Terai, Hirotaka and Yoshikawa, Nobuyuki},
  journal={IEICE Transactions on Electronics},
  volume={105},
  number={6},
  pages={264--269},
  month={June},
  year={2022}
}

@article{przybysz1997interface,
  title={Interface circuits for chip-to-chip data transfer at {GHz} rates},
  author={Przybysz, John X and Miller, DL and Martinet, SS and Kang, Joonhee and Worsham, A Hedge and Farich, ML},
  journal={IEEE Transactions on Applied Superconductivity},
  volume={7},
  number={2},
  pages={2657--2660},
  month={June},
  year={1997}
}

@article{przybysz1999dewar,
  title={Dewar-to-dewar data transfer at {GHz} rates},
  author={Przybysz, JX and McCambridge, JD and Dresselhaus, PD and Worsham, AH and Dean, EJ and Sage, JP and Weir, TJ},
  journal={IEEE Transactions on Applied Superconductivity},
  volume={9},
  number={2},
  pages={2981--2984},
  month={June},
  year={1999}
}

@article{hironaka2025josephson,
  title={Josephson latching driver designed using {10 kA/cm$^2$} {Nb} process as interface for {Josephson-CMOS} hybrid memory},
  author={Hironaka, Yuki and Yoshikawa, Nobuyuki},
  journal={IEEE Transactions on Applied Superconductivity},
  volume={35},
  number={1},
  note = {{A}rt. no. 1300306},
  month={January},
  year={2025}
}

@article{harada2004josephson,
  title={Josephson latching driver with a low bit-error rate},
  author={Harada, Naoki and Yoshikawa, Nobuyuki and Yoshida, Akira and Yokoyama, Naoki},
  journal={IEEE Transactions on Applied Superconductivity},
  volume={14},
  number={4},
  pages={2031--2036},
  month={December},
  year={2004}
}

@article{feng2003josephson,
  title={{Josephson-CMOS} hybrid memory with ultra-high-speed interface circuit},
  author={Feng, YJ and Meng, X and Whiteley, SR and Van Duzer, T and Fujiwara, K and Miyakawa, H and Yoshikawa, N},
  journal={IEEE Transactions on Applied Superconductivity},
  volume={13},
  number={2},
  pages={467--470},
  month={June},
  year={2003}
}

@article{aghighi2018level,
  title={Level shifting circuit for hybrid superconductor-to-semiconductor interface},
  author={Aghighi, F and Jamasb, S and Mazaheri, M},
  journal={Physica C: Superconductivity and its Applications},
  volume={552},
  pages={57--60},
  month={September},
  year={2018},
  publisher={Elsevier}
}

@article{wei2010new,
  title={New {Josephson-CMOS} interface amplifier},
  author={Wei, Daniel and Whiteley, Stephen R and Zheng, Lizhen and Park, Heejoung and Kim, Hoki and Van Duzer, Theodore},
  journal={IEEE Transactions on Applied Superconductivity},
  volume={21},
  number={3},
  pages={805--808},
  month={June},
  year={2010}
}

@article{harada2003logic,
  title={Logic operation at {5 Gb/s} of an output interface for single-flux-quantum systems},
  author={Harada, Naoki and Yoshikawa, Nobuyuki and Yoda, Kenichi and Yoshida, Akira and Yokoyama, Naoki},
  journal={IEEE Transactions on Applied Superconductivity},
  volume={13},
  number={3},
  pages={3814--3816},
  month={September},
  year={2003}
}

@article{nakagawa1982operating,
  title={Operating characteristics of {Josephson} four-junction logic {(4JL)} gate},
  author={Nakagawa, Hiroshi and Sogawa, Eiichi and Kosaka, Shin and Takada, Susumu and Hayakawa, Hisao},
  journal={Japanese Journal of Applied Physics},
  volume={21},
  number={4A},
  pages={L198-L200},
  month={April},
  year={1982},
  publisher={IOP Publishing}
}

@article{takada1980current,
  title={Current injection logic gate with four {Josephson} junctions},
  author={Takada, Susumu and Kosaka, Shin and Hayakawa, Hisao},
  journal={Japanese Journal of Applied Physics},
  volume={19},
  number={S1},
  pages={607-611},
  year={1980},
  publisher={IOP Publishing}
}

@article{hato2005output,
  title={Output interface with latching driver for {LTS-SFQ} circuits},
  author={Hato, Tsunehino and Horibe, Masahiro and Wakana, Hironori and Hidaka, Mutsuo and Tanabe, Keiichi},
  journal={IEEE Transactions on Applied Superconductivity},
  volume={15},
  number={1},
  pages={1--5},
  month={March},
  year={2005}
}

@article{suzuki2009possible,
  title={Possible application of flash-type {SFQ A/D} converter to optical communication systems and their measuring instruments},
  author={Suzuki, Hideo and Maruyama, Michitaka and Hashimoto, Yoshihiro and Fujiwara, Kan and Hidaka, Mutsuo},
  journal={IEEE Transactions on Applied Superconductivity},
  volume={19},
  number={3},
  pages={611--616},
  month={June},
  year={2009}
}

@misc{suzuki2007superconducting,
  title={Superconducting circuit},
  author={Suzuki, Hideo and Tanabe, Keiichi},
  month={September},
  year={2007},
  publisher={Google Patents},
  note={{US} Patent 7,268,713 B2}
}

@article{mustafa2022optimization,
  title={Optimization of {Suzuki} Stack Circuit to Reduce
Power Dissipation},
  author={Y. Mustafa and S. Köse},
  journal={IEEE Transactions on Applied Superconductivity},
  volume={32},
  number={8},
  note = {{A}rt. no. 1301407},
  month={November},
  year={2022}
}

@article{mustafa2023suzuki,
  title={Suzuki Stack Circuit with Differential Output},
  author={{Mustafa}, Yerzhan and K{\"o}se, Sel{\c{c}}uk},
  journal={IEEE Transactions on Applied Superconductivity},
  volume={33},
  number={2},
  note={{A}rt. no. 1300306},
  month={March},
  year={2023}
}

@article{mustafa2024dc,
  title={{DC}-biased {Suzuki} stack circuit for {Josephson-CMOS} memory applications},
  author={Mustafa, Yerzhan and Krause, Keith and Shah, Archit and Hamilton, Michael C and K{\"o}se, Sel{\c{c}}uk},
  journal={Superconductor Science and Technology},
  volume={37},
  number={8},
  note={{A}rt. no. 085023},
  month={July},
  year={2024},
  publisher={IOP Publishing}
}

@article{mustafa2025ternary,
  title={Ternary Digital Output Data Link
from {SFQ} Circuits},
  author={{Mustafa}, Yerzhan and K{\"o}se, Sel{\c{c}}uk},
  journal={IEEE Transactions on Applied Superconductivity},
  volume={35},
  number={5},
  note={{A}rt. no. 1300405},
  month={August},
  year={2025}
}

@article{krause2025signal,
  title={Signal integrity simulations of {4JL} gate pulses from {4 K to 50 K}},
  author={Krause, Keith and Mustafa, Yerzhan and Shah, Archit and K{\"o}se, Sel{\c{c}}uk and Hamilton, Michael C},
  journal={IEEE Transactions on Applied Superconductivity},
  volume={35},
  number={5},
  note = {{A}rt. no. 1300506},
  month={August},
  year={2025}
}

@article{liu2005simulation,
  title={Simulation and measurements on a 64-kbit hybrid {Josephson-CMOS} memory},
  author={Liu, Q and Van Duzer, T and Meng, X and Whiteley, SR and Fujiwara, K and Tomida, T and Tokuda, K and Yoshikawa, N},
  journal={IEEE Transactions on Applied Superconductivity},
  volume={15},
  number={2},
  pages={415--418},
  month={June},
  year={2005}
}

@article{liu2007latency,
  title={Latency and power measurements on a 64-kb hybrid {Josephson-CMOS} memory},
  author={Liu, Q and Fujiwara, K and Meng, X and Whiteley, SR and Van Duzer, T and Yoshikawa, Nobuyuki and Thakahashi, Y and Hikida, T and Kawai, N},
  journal={IEEE Transactions on Applied Superconductivity},
  volume={17},
  number={2},
  pages={526--529},
  month={June},
  year={2007}
}

@article{fujiwara2010new,
  title={New delay-time measurements on a 64-kb {Josephson--CMOS} hybrid memory with a 600-ps access time},
  author={Fujiwara, Kan and Liu, Qingguo and Van Duzer, Theodore and Meng, Xiaofan and Yoshikawa, Nobuyuki},
  journal={IEEE Transactions on Applied Superconductivity},
  volume={20},
  number={1},
  pages={14--20},
  month={February},
  year={2010}
}

@article{kojima2005parameter,
  title={Parameter optimization of a {Josephson} latching driver based on bit-error-rate simulations},
  author={Kojima, H and Yamashiro, Y and Fujiwara, K and Yoshikawa, N and Fujimaki, Akira and Terai, H and Yorozu, S},
  journal={Physica C: Superconductivity and its applications},
  volume={426-431},
  pages={1680--1686},
  month={October},
  year={2005},
  publisher={Elsevier}
}

@article{harada2000multigigahertz,
  title={A multigigahertz {Josephson}-semiconductor interface circuit using {77-K} differential monolithic {HEMT} amplifier and {4.2-K JJ} high-voltage driver for superconductor-semiconductor electronic hybrid systems},
  author={Harada, Naoki and Watanabe, Akira and Awano, Yuji and Hikosaka, Kohki and Yokoyama, Naoki},
  journal={IEEE Journal of Solid-State Circuits},
  volume={35},
  number={1},
  pages={66--73},
  month={January},
  year={2000}
}

@article{sandell1999high,
  title={High data rate switch with amplifier chip},
  author={Sandell, Robert D and Spargo, John W and Leung, Michael and Whiteley, Stephen R},
  journal={IEEE Transactions on Applied Superconductivity},
  volume={9},
  number={2},
  pages={2985--2988},
  month={June},
  year={1999}
}

@article{van2002hybrid,
  title={Hybrid {Josephson-CMOS} memory: a solution for the {Josephson} memory problem},
  author={Van Duzer, Theodore and Feng, Yijun and Meng, Xiaofan and Whiteley, Stephen R and Yoshikawa, Nobuyuki},
  journal={Superconductor Science and Technology},
  volume={15},
  number={12},
  pages={1669-1674},
  month={November},
  year={2002},
  publisher={IOP Publishing}
}

@article{hato2003sfq,
  title={{SFQ}-to-level logic conversion by {HTS Josephson} drivers for output interface},
  author={Hato, Tsunehiro and Ishimaru, Yoshihiro and Harada, Naoki and Horibe, Masahiro and Yoshida, Akira and Tarutani, Yoshinobu and Tanabe, Keiichi and Yokoyama, Naoki},
  journal={IEEE Transactions on Applied Superconductivity},
  volume={13},
  number={2},
  pages={397--400},
  month={June},
  year={2003}
}

@article{zhao2025high,
  title={A high voltage and high speed superconductive voltage driver using a damped asymmetric {DC SQUID} array},
  author={Zhao, Mengfei and Wang, Yongliang and Gao, Xiaoping and Yuan, Pusheng and Wang, Shuna and Niu, Minghui and You, Lixing and Ren, Jie and Li, Lingyun},
  journal={Superconductor Science and Technology},
  volume={38},
  number={4},
  note = {{A}rt. no. 045011},
  month={March},
  year={2025},
  publisher={IOP Publishing}
}

@article{herr2010high,
  title={A high-efficiency superconductor distributed amplifier},
  author={Herr, Quentin P},
  journal={Superconductor Science and Technology},
  volume={23},
  number={2},
  note = {{A}rt. no. 022004},
  month={January},
  year={2010},
  publisher={IOP Publishing}
}

@article{zhao2025high_magnetometer,
  title={A high-voltage {SFQ-to-DC} driver for wide-range digital {SQUID} magnetometer based on flux quanta counting scheme},
  author={Zhao, Mengfei and Wang, Yongliang and Yuan, Pusheng and Wang, Shuna and Li, Lingyun and You, Lixing},
  journal={Physica C: Superconductivity and its Applications},
  volume={633},
  note = {{A}rt. no. 1354708},
  month={June},
  year={2025},
  publisher={Elsevier}
}

@article{koch1999nrz,
  title={A {NRZ}-output amplifier for {RSFQ} circuits},
  author={Koch, R and Ostertag, P and Crocoll, E and Gotz, M and Neuhaus, M and Scherer, T and Winter, M and Jutzi, W},
  journal={IEEE Transactions on Applied Superconductivity},
  volume={9},
  number={2},
  pages={3549--3552},
  month={June},
  year={1999}
}

@article{welty1991series,
  title={A series array of {DC SQUIDs}},
  author={Welty, Richard P and Martinis, John M},
  journal={IEEE Transactions on Magnetics},
  volume={27},
  number={2},
  pages={2924--2926},
  month={March},
  year={1991}
}

@article{yoshikawa2001component,
  title={Component development for a 16 {Gb/s RSFQ-CMOS} interface system},
  author={Yoshikawa, Nobuyuki and Abe, Takashi and Kato, Yohsuke and Hoshina, Hiroshi},
  journal={IEEE Transactions on Applied Superconductivity},
  volume={11},
  number={1},
  pages={735--738},
  month={March},
  year={2001}
}

@inproceedings{higuchi2019design,
  title={Design and operation of distributed double-{SQUID} amplifier for {RSFQ} circuits},
  author={Higuchi, Komei and Shimada, Hiroshi and Mizugaki, Yoshinao},
  booktitle={Journal of Physics: Conference Series},
  volume={1293},
  number={1},
  note = {{A}rt. no. 012060},
  year={2019},
  organization={IOP Publishing}
}

@article{gupta2019digital,
  title={Digital output data links from superconductor integrated circuits},
  author={Gupta, Deepnarayan and Sarwana, Saad and Kirichenko, Dmitri and Dotsenko, Vladimir and Lehmann, A Erik and Filippov, Timur V and Wong, Wei-Ting and Chang, Su-Wei and Ravindran, Prasanna and Bardin, Joseph},
  journal={IEEE Transactions on Applied Superconductivity},
  volume={29},
  number={5},
  note = {{A}rt. no. 1303208},
  month={August},
  year={2019}
}

@article{mizugaki2019enhanced,
  title={Enhanced voltage swing of rapid-single-flux-quantum distributed output amplifier equipped with double-stack superconducting quantum interference devices},
  author={Mizugaki, Yoshinao and Higuchi, Komei and Shimada, Hiroshi},
  journal={IEICE Electronics Express},
  volume={16},
  number={14},
  pages={1--4},
  month={August},
  year={2019}
}

@article{soloviev2007high,
  title={High voltage driver for {RSFQ} digital signal processor},
  author={Soloviev, Igor I and Rafique, M Raihan and Engseth, Henrik and Kidiyarova-Shevchenko, Anna},
  journal={IEEE Transactions on Applied Superconductivity},
  volume={17},
  number={2},
  pages={470--473},
  month={June},
  year={2007}
}

@article{hashimoto2007implementation,
  title={Implementation and experimental evaluation of a cryocooled system prototype for high-throughput {SFQ} digital applications},
  author={Hashimoto, Yoshihito and Yorozu, Shinichi and Miyazaki, Toshiyuki and Kameda, Yoshio and Suzuki, Hideo and Yoshikawa, Nobuyuki},
  journal={IEEE Transactions on Applied Superconductivity},
  volume={17},
  number={2},
  pages={546--551},
  month={June},
  year={2007}
}

@article{herr2007inductive,
  title={Inductive isolation in stacked {SQUID} amplifiers},
  author={Herr, Quentin P and Miller, Donald L and Pesetski, Aaron A and Przybysz, John X},
  journal={IEEE Transactions on Applied Superconductivity},
  volume={17},
  number={2},
  pages={565--568},
  month={June},
  year={2007}
}

@article{mukhanov1997josephson,
  title={Josephson output interfaces for {RSFQ} circuits},
  author={Mukhanov, OA and Rylov, Sergey V and Gaidarenko, Dmitri V and Dubash, Noshir B and Borzenets, Valery V},
  journal={IEEE Transactions on Applied Superconductivity},
  volume={7},
  number={2},
  pages={2826--2831},
  month={June},
  year={1997}
}

@article{terai2009readout,
  title={Readout electronics using single-flux-quantum circuit technology for superconducting single-photon detector array},
  author={Terai, Hirotaka and Miki, Shigetoshi and Wang, Zhen},
  journal={IEEE Transactions on Applied Superconductivity},
  volume={19},
  number={3},
  pages={350--353},
  month={June},
  year={2009}
}

@article{hashimoto2009measurement,
  title={Measurement of Superconductive Voltage Drivers up to 25 {Gb/s/ch}},
  author={Hashimoto, Yoshihito and Suzuki, Hideo and Nagasawa, Shuichi and Maruyama, Michitaka and Fujiwara, Kan and Hidaka, Mutsuo},
  journal={IEEE Transactions on Applied Superconductivity},
  volume={19},
  number={3},
  pages={1022--1025},
  month={June},
  year={2009}
}

@article{razmkhah2021compact,
  title={A compact high frequency voltage amplifier for superconductor--semiconductor logic interface},
  author={Razmkhah, Sasan and Bozbey, Ali and Febvre, Pascal},
  journal={Superconductor Science and Technology},
  volume={34},
  number={4},
  note = {{A}rt. no. 045013},
  month={February},
  year={2021},
  publisher={IOP Publishing}
}

@inproceedings{mustafa2025spam,
  title={{S-PAM}: Superconductor-Semiconductor Interface Circuit with Pulse-Amplitude Modulation},
  author={Mustafa, Yerzhan and K{\"o}se, Sel{\c{c}}uk},
  booktitle={IEEE International Symposium on Circuits and Systems (ISCAS)},
  pages={1--5},
  month={May},
  year={2025}
}

@article{castellanos2021single,
  title={Single-flux-quantum multiplier circuits for synthesizing gigahertz waveforms with quantum-based accuracy},
  author={Castellanos-Beltran, Manuel A and Olaya, DI and Sirois, AJ and Donnelly, CA and Dresselhaus, PD and Benz, SP and Hopkins, PF},
  journal={IEEE Transactions on Applied Superconductivity},
  volume={31},
  number={3},
  note = {{A}rt. no. 1400109},
  month={April},
  year={2021}
}

@inproceedings{khabipov2006development,
  title={Development of {RSFQ} voltage drivers for arbitrary {AC} waveform synthesisers},
  author={Khabipov, MI and Hagedorn, D and Buchholz, F-Im and Kohlmann, J and Maibaum, F and Schilling, M and Niemeyer, J},
  booktitle={Journal of Physics: Conference Series},
  volume={43},
  number={1},
  pages={1175--1178},
  year={2006},
  organization={IOP Publishing}
}

@article{hirayama2002characteristics,
  title={Characteristics of a voltage multiplier for a {RSFQ} digital-to-analog converter},
  author={Hirayama, Fuminori and Maezawa, Masaaki and Kiryu, Shogo and Sasaki, Hitoshi and Shoji, Akira},
  journal={Superconductor Science and Technology},
  volume={15},
  number={4},
  pages={494--498},
  month={February},
  year={2002},
  publisher={IOP Publishing}
}

@article{hirayama2003characteristics,
  title={Characteristics of voltage multipliers for a {Josephson D/A} converter},
  author={Hirayama, Fuminori and Maezawa, Masaaki and Sasaki, Hitoshi and Shoji, Akira},
  journal={IEEE Transactions on Applied Superconductivity},
  volume={13},
  number={2},
  pages={484--487},
  month={June},
  year={2003}
}

@article{inamdar2009superconducting,
  title={Superconducting switching amplifiers for high speed digital data links},
  author={Inamdar, Amol and Rylov, Sergey and Sarwana, Saad and Gupta, Deepnarayan},
  journal={IEEE Transactions on Applied Superconductivity},
  volume={19},
  number={3},
  pages={1026--1033},
  month={June},
  year={2009}
}

@article{dubash2000system,
  title={System demonstration of a multigigabit network switch},
  author={Dubash, Noshir B and Borzenets, Valery V and Zhang, Yongming M and Kaplunenko, Vsevolod and Spargo, John W and Smith, AD and Van Duzer, Theodore},
  journal={IEEE Transactions on Microwave Theory and Techniques},
  volume={48},
  number={7},
  pages={1209--1215},
  month={July},
  year={2000}
}

@article{egan2022true,
  title={True differential superconducting on-chip output amplifier},
  author={Egan, Jonathan and Brownfield, Andrew and Herr, Quentin},
  journal={Superconductor Science and Technology},
  volume={35},
  number={4},
  note = {{A}rt. no. 045018},
  month={March},
  year={2022},
  publisher={IOP Publishing}
}

@article{tarutani2001interface,
  title={Interface circuit using {JTLs} as control lines of {SQUID} array},
  author={Tarutani, Yoshinobu and Saitoh, Kazuo and Takagi, Kazumasa},
  journal={IEEE Transactions on Applied Superconductivity},
  volume={11},
  number={1},
  pages={341--344},
  month={March},
  year={2001}
}

@article{kornev2007development,
  title={Development of {SQIF}-based output broad band amplifier},
  author={Kornev, Victor K and Soloviev, Igor I and Klenov, Nikolai V and Mukhanov, Oleg A},
  journal={IEEE Transactions on Applied Superconductivity},
  volume={17},
  number={2},
  pages={569--572},
  month={June},
  year={2007}
}

@article{kaplunenko1989experimental,
  title={Experimental study of the {RSFQ} logic elements},
  author={Kaplunenko, VK and Khabipov, MI and Koshelets, VP and Likharev, KK and Mukhanov, OA and Semenov, VK and Serpuchenko, IL and Vystavkin, AN},
  journal={IEEE Transactions on Magnetics},
  volume={25},
  number={2},
  pages={861--864},
  month={March},
  year={1989}
}

@article{ortlepp2009superconductor,
  title={Superconductor-to-semiconductor interface circuit for high data rates},
  author={Ortlepp, Thomas and Wuensch, Stefan and Schubert, Marco and Febvre, Pascal and Ebert, Bjoern and Kunert, Juergen and Crocoll, Erich and Meyer, Hans-Georg and Siegel, Michael and Uhlmann, F Hermann},
  journal={IEEE Transactions on Applied Superconductivity},
  volume={19},
  number={1},
  pages={28--34},
  month={February},
  year={2009}
}

@misc{sunysb_rsfq_cell_library, 
title={{Stony Brook University}: Rapid Single-Flux-Quantum Laboratory}, 
url={https://you.stonybrook.edu/rsfq/lib/}, 
publisher={Stony Brook University},
note={{A}ccessed: July 26, 2025}}

@misc{supertools_rsfq_cell_library, 
title={{SuperTools/ColdFlux RSFQ} Cell Library}, 
url={https://github.com/sunmagnetics/RSFQlib}, 
publisher={GitHub},
note={{A}ccessed: July 26, 2025}}

@article{tanaka2025low,
  title={Low-Power Single-Flux-Quantum Standard Cell Library Using 250 {A}/cm $\^{}$\{$2$\}$ $ Process for Qubit Control Applications},
  author={Tanaka, Masamitsu and Kitagawa, Yoshihiro and Satoh, Tetsuro and Yamamoto, Tsuyoshi},
  journal={IEEE Transactions on Applied Superconductivity},
  volume={35},
  number={5},
  note = {{A}rt. no. 1700405},
  month={August},
  year={2025}
}

@article{maezawa2004design,
  title={Design and fabrication of {RSFQ} cell library for middle-scale applications},
  author={Maezawa, Masaaki and Hirayama, Fuminori and Suzuki, Motohiro},
  journal={Physica C: Superconductivity},
  volume={412-414},
  pages={1591--1596},
  month={October},
  year={2004},
  publisher={Elsevier}
}

@article{cong2024superconductor,
  title={Superconductor logic implementation with {all-JJ} inductor-free cell library},
  author={Cong, Haolin and Razmkhah, Sasan and Karamuftuoglu, Mustafa Altay and Pedram, Massoud},
  journal={IEEE Transactions on Applied Superconductivity},
  volume={34},
  number={9},
  note = {{A}rt. no. 1301110},
  month={December},
  year={2024}
}

@article{chen2025signal,
  title={Signal Matching from {SFQ/DC} Converter to {SiGe BiCMOS} Interface},
  author={Chen, Zhichao and Li, Lingyun and You, Lixing},
  journal={IEICE Electronics Express},
  pages={1--6},
  volume={21},
  number={1},
  month={May},
  year={2025}
}

@article{zhao2017nanocryotron,
  title={A nanocryotron comparator can connect single-flux-quantum circuits to conventional electronics},
  author={Zhao, Qing-Yuan and McCaughan, Adam N and Dane, Andrew E and Berggren, Karl K and Ortlepp, Thomas},
  journal={Superconductor Science and Technology},
  volume={30},
  number={4},
  note = {{A}rt. no. 044002},
  month={March},
  year={2017},
  publisher={IOP Publishing}
}

@article{mccaughan2019superconducting,
  title={A superconducting thermal switch with ultrahigh impedance for interfacing superconductors to semiconductors},
  author={McCaughan, Adam N and Verma, Varun B and Buckley, Sonia M and Allmaras, JP and Kozorezov, AG and Tait, AN and Nam, SW and Shainline, JM},
  journal={Nature Electronics},
  volume={2},
  number={10},
  pages={451--456},
  month={September},
  year={2019}
}

@article{mccaughan2014superconducting,
  title={A superconducting-nanowire three-terminal electrothermal device},
  author={McCaughan, Adam N and Berggren, Karl K},
  journal={Nano letters},
  volume={14},
  number={10},
  pages={5748--5753},
  month={September},
  year={2014},
  publisher={ACS Publications}
}

@article{tanaka2017josephson,
  title={{Josephson-CMOS} hybrid memory with nanocryotrons},
  author={Tanaka, Masamitsu and Suzuki, Masato and Konno, Gen and Ito, Yuki and Fujimaki, Akira and Yoshikawa, Nobuyuki},
  journal={IEEE Transactions on Applied Superconductivity},
  volume={27},
  number={4},
  note = {{A}rt. no. 1800904},
  month={June},
  year={2017}
}

@article{sano2018thermally,
  title={Thermally assisted superconductor transistors for {Josephson-CMOS} hybrid memories},
  author={Sano, Kyosuke and Suzuki, Masato and Maruyama, Kohei and Taniguchi, Soya and Tanaka, Masamitsu and Fujimaki, Akira and Inoue, Masumi and Yoshikawa, Nobuyuki},
  journal={IEICE Transactions on Electronics},
  volume={101},
  number={5},
  pages={370--377},
  month={May},
  year={2018}
}

@article{nguyen2020cryogenic,
  title={Cryogenic memory architecture integrating spin {Hall} effect based magnetic memory and superconductive cryotron devices},
  author={Nguyen, Minh-Hai and Ribeill, Guilhem J and Gustafsson, Martin V and Shi, Shengjie and Aradhya, Sriharsha V and Wagner, Andrew P and Ranzani, Leonardo M and Zhu, Lijun and Baghdadi, Reza and Butters, Brenden and others},
  journal={Scientific Reports},
  volume={10},
  number={1},
  note = {{A}rt. no. 248},
  month={January},
  year={2020}
}

@article{zhao2018compact,
  title={A compact superconducting nanowire memory element operated by nanowire cryotrons},
  author={Zhao, Qing-Yuan and Toomey, Emily A and Butters, Brenden A and McCaughan, Adam N and Dane, Andrew E and Nam, Sae-Woo and Berggren, Karl K},
  journal={Superconductor Science and Technology},
  volume={31},
  number={3},
  note = {{A}rt. no. 035009},
  month={February},
  year={2018},
  publisher={IOP Publishing}
}

@article{alam2023cryogenic,
  title={Cryogenic reconfigurable logic with superconducting heater cryotron: Enhancing area efficiency and enabling camouflaged processors},
  author={Alam, Shamiul and Rampini, Dana S and Oripov, Bakhrom G and McCaughan, Adam N and Aziz, Ahmedullah},
  journal={Applied Physics Letters},
  volume={123},
  number={15},
  note = {{A}rt. no. 152603},
  month={October},
  year={2023},
  publisher={AIP Publishing}
}

@article{baghdadi2020multilayered,
  title={Multilayered heater nanocryotron: A superconducting-nanowire-based thermal switch},
  author={Baghdadi, Reza and Allmaras, Jason P and Butters, Brenden A and Dane, Andrew E and Iqbal, Saleem and McCaughan, Adam N and Toomey, Emily A and Zhao, Qing-Yuan and Kozorezov, Alexander G and Berggren, Karl K},
  journal={Physical Review Applied},
  volume={14},
  number={5},
  note = {{A}rt. no. 054011},
  month={November},
  year={2020},
  publisher={APS}
}

@article{paul2025photolithography,
  title={Photolithography-compatible three-terminal superconducting switch for driving {CMOS} loads},
  author={Paul, Dip Joti and Zhou, Tony X and Berggren, Karl K},
  journal={Physical Review Applied},
  volume={24},
  number={2},
  month={August},
  note = {{A}rt. no. 024060},
  year={2025},
  publisher={APS}
}

@article{nevirkovets2020characterization,
  title={Characterization of amplification properties of the superconducting-ferromagnetic transistor},
  author={Nevirkovets, Ivan P and Kojima, Takafumi and Uzawa, Yoshinori and Kotula, Paul G and Missert, Nancy and Mukhanov, Oleg A},
  journal={IEEE Transactions on Applied Superconductivity},
  volume={30},
  number={7},
  note = {{A}rt. no. 1800105},
  month={October},
  year={2020}
}

@article{hebard1979dc,
  title={A dc-powered {Josephson} flip-flop},
  author={Hebard, A and Pei, S and Dunkleberger, L and Fulton, T},
  journal={IEEE Transactions on Magnetics},
  volume={15},
  number={1},
  pages={408--411},
  month={January},
  year={1979}
}

@article{schneider1995broadband,
  title={Broadband interfacing of superconducting digital systems to room temperature electronics},
  author={Schneider, DF and Lin, JC and Polonsky, SV and Semenov, VK and Hamilton, CA},
  journal={IEEE Transactions on Applied Superconductivity},
  volume={5},
  number={2},
  pages={3152--3155},
  month={June},
  year={1995}
}

@article{polonsky1991new,
  title={New {SFQ/DC} converter for {RSFQ} logic/memory family},
  author={Polonsky, SV},
  journal={Superconductor Science and Technology},
  volume={4},
  number={9},
  pages={442--444},
  year={1991},
  publisher={IOP Publishing}
}

@article{hatano1992performance,
  title={Performance analysis of the {Josephson} dc flip-flop},
  author={Hatano, Yuji and Nagaishi, Hideyuki and Nakahara, Kouji and Kawabe, Ushio},
  journal={IEEE Transactions on Applied Superconductivity},
  volume={2},
  number={3},
  pages={148--155},
  month={September},
  year={1992}
}

@article{kotera1983ring,
  title={Ring oscillator experiment using a huffle circuit},
  author={Kotera, N and Asano, A and Harada, Y and Kawabe, U},
  journal={IEEE Transactions on Magnetics},
  volume={19},
  number={3},
  pages={1174--1177},
  month={May},
  year={1983}
}

@article{herr2005stacked,
  title={Stacked double-flux-quantum output amplifier},
  author={Herr, Quentin P},
  journal={IEEE Transactions on Applied Superconductivity},
  volume={15},
  number={2},
  pages={259--262},
  month={June},
  year={2005}
}

@article{golomidov1992single,
  title={Single flux quantum voltage amplifiers},
  author={Golomidov, Vladimir and Kaplunenko, Vsevolod and Khabipov, Marat and Koshelets, Valery and Kaplunenko, Olga},
  journal={Cryogenics},
  volume={32},
  pages={509--512},
  year={1992},
  publisher={Elsevier}
}

@article{kaplunenko1991single,
  title={Single flux quantum quasi-digital voltage amplifier},
  author={Kaplunenko, VK and Koshelets, VP and Khabipov, MI and Golomidov, VM and Kovtonyuk, SA},
  journal={Superconductor Science and Technology},
  volume={4},
  number={11},
  pages={671--673},
  year={1991},
  publisher={IOP Publishing}
}

@inproceedings{mukhanov2013development,
  title={Development of Energy-efficient Cryogenic Optical {(ECO)} data link},
  author={Mukhanov, OA and Vernik, I and Kirichenko, A and Kadin, A and Choquette, Kent D and Tan, MP and Fryslie, T},
  booktitle={IEEE 14th International Superconductive Electronics Conference (ISEC)},
  pages={1--3},
  month={July},
  year={2013}
}

@article{sakai2020proposal,
  title={Proposal of ultra-low voltage quantum well optical modulator for optical interconnection in superconducting integrated circuit systems},
  author={Sakai, Kota and Kato, Seiji and Yoshikawa, Nobuyuki and Kokubun, Yasuo and Arakawa, Taro},
  journal={Japanese Journal of Applied Physics},
  volume={59},
  number={SO},
  note = {{A}rt. no. SOOB01},
  month={April},
  year={2020},
  publisher={IOP Publishing}
}

@article{likharev1991rsfq,
  title={{RSFQ} logic/memory family: A new {Josephson-junction} technology for sub-terahertz-clock-frequency digital systems},
  author={Likharev, Konstantin K and Semenov, Vasilii K},
  journal={IEEE Transactions on Applied Superconductivity},
  volume={1},
  number={1},
  pages={3--28},
  month={March},
  year={1991}
}

@article{morooka1997design,
  title={Design, fabrication and evaluation of a four-Josephson-junction superconducting quantum interference device with two superconducting loops},
  author={Morooka, Toshimitsu Morooka Toshimitsu},
  journal={Japanese Journal of Applied Physics},
  volume={36},
  number={12A},
  pages={L1587--L1590},
  month={December},
  year={1997},
  publisher={IOP Publishing}
}

@article{wuensch2009cryogenic,
  title={Cryogenic semiconductor amplifier for {RSFQ}-circuits with high data rates at 4.2 {K}},
  author={Wuensch, Stefan and Ortlepp, Thomas and Crocoll, Erich and Uhlmann, Friedrich Hermann and Siegel, Michael},
  journal={IEEE Transactions on Applied Superconductivity},
  volume={19},
  number={3},
  pages={574--579},
  month={June},
  year={2009}
}

@article{somei2021enhanced,
  title={Enhanced operation frequencies of bipolar double-flux-quantum amplifiers fabricated using 10 {kA} cm$^{-2}$ {Nb/AlOx/Nb} integration process},
  author={Somei, Yuta and Shimada, Hiroshi and Mizugaki, Yoshinao},
  journal={Japanese Journal of Applied Physics},
  volume={60},
  number={7},
  note={{A}rt. no. 073001},
  month={July},
  year={2021},
  publisher={IOP Publishing}
}

@article{mizugaki20191000,
  title={1000-fold double-flux-quantum voltage multiplier employing directional propagation of flux quanta through asymmetrically damped junction branches},
  author={Mizugaki, Yoshinao and Arai, Yuma and Watanabe, Tomoki and Shimada, Hiroshi},
  journal={IEEE Transactions on Applied Superconductivity},
  volume={29},
  number={5},
  note={{A}rt. no. 1400105},
  month={August},
  year={2019}
}

@article{yasukawa2024ginzburg,
  title={{Ginzburg--Landau} simulations of three-terminal operation of a superconducting nanowire cryotron},
  author={Yasukawa, Naoki and Nishio, Taichiro and Mawatari, Yasunori},
  journal={Superconductor Science and Technology},
  volume={37},
  number={6},
  note={{A}rt. no. 065013},
  month={May},
  year={2024},
  publisher={IOP Publishing}
}

@article{karam2025parameter,
  title={Parameter extraction for a {SPICE} model of an {hTron} superconducting thermal switch},
  author={Karam, Valentin and Medeiros, Owen and Dandachi, Tareq El and Castellani, Matteo and Foster, Reed and Berggren, Karl and Colangelo, Marco},
  journal={Physical Review Applied},
  volume={24},
  number={2},
  note={{A}rt. no. 024020},
  month={August},
  year={2025},
  publisher={APS}
}

@article{buck2007cryotron,
  title={The cryotron-a superconductive computer component},
  author={Buck, Dudley A},
  journal={Proceedings of the IRE},
  volume={44},
  number={4},
  pages={482--493},
  month={April},
  year={1956},
  publisher={IEEE}
}

@article{shainline2018circuit,
  title={Circuit designs for superconducting optoelectronic loop neurons},
  author={Shainline, Jeffrey M and Buckley, Sonia M and McCaughan, Adam N and Chiles, Jeff and Jafari-Salim, Amir and Mirin, Richard P and Nam, Sae Woo},
  journal={Journal of Applied Physics},
  volume={124},
  number={15},
  note={{A}rt. no. 152130},
  month={October},
  year={2018},
  publisher={AIP Publishing}
}

@article{faris2003quiteron,
  title={Quiteron},
  author={Faris, S and Raider, S and Gallagher, W and Drake, R},
  journal={IEEE Transactions on Magnetics},
  volume={19},
  number={3},
  pages={1293--1295},
  month={May},
  year={2003}
}

@article{frank1984circuit,
  title={A circuit-oriented quiteron analysis},
  author={Frank, David J},
  journal={Journal of Applied Physics},
  volume={56},
  number={9},
  pages={2553--2557},
  month={May},
  year={1984},
  publisher={American Institute of Physics}
}

@article{nevirkovets2009superconducting,
  title={A superconducting transistorlike device having good input-output isolation.},
  author={Nevirkovets, IP},
  journal={Applied Physics Letters},
  volume={95},
  number={5},
  month={August},
  note={{A}rt. no. 052505},
  year={2009},
  publisher={AIP Publishing}
}

@article{nevirkovets2014superconducting,
  title={Superconducting-ferromagnetic transistor},
  author={Nevirkovets, Ivan P and Chernyashevskyy, Oleksandr and Prokopenko, Georgy V and Mukhanov, Oleg A and Ketterson, John B},
  journal={IEEE Transactions on Applied Superconductivity},
  volume={24},
  number={4},
  note={{A}rt. no. 1800506},
  month={August},
  year={2014}
}

@article{nevirkovets2023electrically,
  title={Electrically controlled hybrid superconductor--ferromagnet cell for high density cryogenic memory},
  author={Nevirkovets, IP and Mukhanov, OA},
  journal={Applied Physics Letters},
  volume={123},
  number={7},
  note={{A}rt. no. 072601},
  month={August},
  year={2023},
  publisher={AIP Publishing}
}

@article{de2019josephson,
  title={Josephson field-effect transistors based on all-metallic {Al/Cu/Al} proximity nanojunctions},
  author={De Simoni, Giorgio and Paolucci, Federico and Puglia, Claudio and Giazotto, Francesco},
  journal={ACS nano},
  volume={13},
  number={7},
  pages={7871--7876},
  month={June},
  year={2019},
  publisher={ACS Publications}
}

@article{wen2019josephson,
  title={Josephson junction field-effect transistors for Boolean logic cryogenic applications},
  author={Wen, Feng and Shabani, Javad and Tutuc, Emanuel},
  journal={IEEE Transactions on Electron Devices},
  volume={66},
  number={12},
  pages={5367--5374},
  month={December},
  year={2019}
}

@article{akazaki1996josephson,
  title={A {Josephson} field effect transistor using an InAs-inserted-channel {In}$_{0.52}${Al}$_{0. 48}${As}/{In}$_{0. 53}{Ga}$_{0. 47}${As} inverted modulation-doped structure},
  author={Akazaki, Tatsushi and Takayanagi, Hideaki and Nitta, Junsaku and Enoki, Takatomo},
  journal={Applied Physics Letters},
  volume={68},
  number={3},
  pages={418--420},
  month={January},
  year={1996},
  publisher={American Institute of Physics}
}

@article{nishino1989sufet,
  title={0.1-mu m gate-length superconducting {FET}},
  author={Nishino, T and Hatano, M and Hasegawa, H and Murai, F and Kure, T and Hiraiwa, A and Yagi, K and Kawabe, Ushio},
  journal={IEEE Electron Device Letters},
  volume={10},
  number={2},
  pages={61--63},
  month={February},
  year={1989}
}

@article{paolucci2018ultra,
  title={Ultra-efficient superconducting {Dayem} bridge field-effect transistor},
  author={Paolucci, Federico and De Simoni, Giorgio and Strambini, Elia and Solinas, Paolo and Giazotto, Francesco},
  journal={Nano Letters},
  volume={18},
  number={7},
  pages={4195--4199},
  month={June},
  year={2018},
  publisher={ACS Publications}
}

@article{de2018metallic,
  title={Metallic supercurrent field-effect transistor},
  author={De Simoni, Giorgio and Paolucci, Federico and Solinas, Paolo and Strambini, Elia and Giazotto, Francesco},
  journal={Nature Nanotechnology},
  volume={13},
  number={9},
  pages={802--805},
  month={September},
  year={2018},
  publisher={Nature Publishing Group UK London}
}

@article{giazotto2010superconducting,
  title={Superconducting quantum interference proximity transistor},
  author={Giazotto, Francesco and Peltonen, Joonas T and Meschke, Matthias and Pekola, Jukka P},
  journal={Nature Physics},
  volume={6},
  number={4},
  pages={254--259},
  month={April},
  year={2010},
  publisher={Nature Publishing Group UK London}
}

@article{yohannes2023high,
  title={High density fabrication process for single flux quantum circuits},
  author={Yohannes, D and Renzullo, M and Vivalda, J and Jacobs, AC and Yu, M and Walter, J and Kirichenko, AF and Vernik, IV and Mukhanov, OA},
  journal={Applied Physics Letters},
  volume={122},
  number={21},
  month={May},
  note={{A}rt. no. 212601},
  year={2023},
  publisher={AIP Publishing}
}

@phdthesis{liu2007josephson,
  title={{Josephson-CMOS} hybrid memories},
  author={Liu, Qingguo},
  year={2007},
  school={University of California, Berkeley}
}

@article{gupta2013low,
  title={Low-power high-speed hybrid temperature heterogeneous technology digital data link},
  author={Gupta, Deepnarayan and Bardin, Joseph C and Inamdar, Amol and Dayalu, Aniruddha and Sarwana, Saad and Ravindran, Prasana and Chang, Su-Wei and Coskun, Ahmet H and Sadrabadi, Mohammad Ghadiri},
  journal={IEEE Transactions on Applied Superconductivity},
  volume={23},
  number={3},
  note={{A}rt. no. 1701806},
  month={June},
  year={2013}
}

@article{ravindran2014power,
  title={Power-optimized temperature-distributed digital data link},
  author={Ravindran, Prasana and Chang, Su-Wei and Gupta, Deepnarayan and Inamdar, Amol and Dotsenko, Vladimir and Sarwana, Saad M and Bardin, Joseph C},
  journal={IEEE Transactions on Applied Superconductivity},
  volume={25},
  number={3},
  note={{A}rt. no. 1300605},
  month={June},
  year={2014}
}

@article{ravindran2017energy,
  title={Energy efficient digital data link},
  author={Ravindran, Prasana and Chang, Su-Wei and Wong, Wei-Ting and Sarwana, Saad M and Dotsenko, Vladimir and Tang, Jia and Ruotolo, Steven and Gupta, Deepnarayan and Bardin, Joseph C},
  journal={IEEE Transactions on Applied Superconductivity},
  volume={27},
  number={4},
  note={{A}rt. no. 1301105},
  month={June},
  year={2017}
}

@article{chen2025sfq,
  title={A {SFQ-to-CMOS} Interface Circuit Based on {SiGe BiCMOS} for {Josephson-CMOS} Hybrid System},
  author={Chen, Zhichao and Zhang, Xingyu and You, Lixing and Li, Lingyun},
  journal={Journal of Low Temperature Physics},
  pages={196--208},
  volume={219},
  month={April},
  year={2025},
  publisher={Springer}
}

@article{kuwabara2013design,
  title={Design and implementation of 64-kb {CMOS} static {RAMs} for {Josephson-CMOS} hybrid memories},
  author={Kuwabara, Keita and Jin, Hyunjoo and Yamanashi, Yuki and Yoshikawa, Nobuyuki},
  journal={IEEE Transactions on Applied Superconductivity},
  volume={23},
  number={3},
  note={{A}rt. no. 1700704},
  month={June},
  year={2013}
}

@article{jin2012investigation,
  title={Investigation of robust {CMOS} amplifiers for {Josephson-CMOS} hybrid memories},
  author={Jin, Hyunjoo and Kuwabara, Keita and Yamanashi, Yuki and Yoshikaw, Nobuyuki},
  journal={Physics Procedia},
  volume={36},
  pages={229--234},
  month={September},
  year={2012},
  publisher={Elsevier}
}

@book{salman2012high,
  title={High performance integrated circuit design},
  author={Salman, Emre and Friedman, Eby G},
  publisher={McGraw Hill Professional},
  year={2012}
}

@book{krylov2024single,
  title={Single Flux Quantum Integrated Circuit Design, Second Edition},
  author={Krylov, Gleb and  Jabbari, Tahereh and Friedman, Eby G.},
  year={2024},
  publisher={Springer}
}

@article{ghoshal1993superconductor,
  title={Superconductor-semiconductor memories},
  author={Ghoshal, Uttam and Kroger, H and Van Duzer, T},
  journal={IEEE Transactions on Applied Superconductivity},
  volume={3},
  number={1},
  pages={2315--2318},
  month={March},
  year={1993}
}

@article{ghoshal1995cmos,
  title={{CMOS} amplifier designs for {Josephson-CMOS} interface circuits},
  author={Ghoshal, Uttam and Kishore, SV and Feldman, AR and Huynh, Luong and Van Duzer, T},
  journal={IEEE Transactions on Applied Superconductivity},
  volume={5},
  number={2},
  pages={2640--2643},
  month={June},
  year={1995}
}

@article{wu2021vcsel,
  title={{2.6 K VCSEL} data link for cryogenic computing},
  author={Wu, Haonan and Fu, Wenning and Feng, Milton and Deppe, Dennis},
  journal={Applied Physics Letters},
  volume={119},
  number={4},
  note={{A}rt. no. 041101},
  month={July},
  year={2021},
  publisher={AIP Publishing}
}

@article{wu2024cryo,
  title={{Cryo-VCSELs} Operated at {2.8 K and 40 K} with Record Bandwidth, Power and Linearity for Optical Data Links in Quantum Computing},
  author={Wu, Haonan and Fu, Wenning and Liu, Zetai and Chaw, Derek and He, Yulin and Feng, Milton},
  journal={IEEE Journal of Quantum Electronics},
  volume={60},
  number={5},
  month={October},
  note={{A}rt. no. 2400310},
  year={2024}
}

@article{namvar2024improving,
  title={Improving p-doped {DBRs} operation at cryogenic temperatures: Investigating different mirror geometry},
  author={Namvar, Behzad and Uusitalo, Topi and Virtanen, Heikki and Guina, Mircea and Viheri{\"a}l{\"a}, Jukka},
  journal={IEEE Photonics Journal},
  volume={16},
  number={4},
  note={{A}rt. no. 1502009},
  month={August},
  year={2024}
}

@article{namvar2025thermal,
  title={Thermal characteristics of a double intra-cavity contact {VCSEL} for cryogenic optical links},
  author={Namvar, Behzad and Rajala, Patrik and Guina, Mircea and Hakkarainen, Teemu and Virtanen, Heikki and Uusitalo, Topi and Viheri{\"a}l{\"a}, Jukka},
  journal={IEEE Photonics Journal},
  volume={17},
  number={5},
  note={{A}rt. no. 1502406},
  month={October},
  year={2025}
}

@article{youssefi2021cryogenic,
  title={A cryogenic electro-optic interconnect for superconducting devices},
  author={Youssefi, Amir and Shomroni, Itay and Joshi, Yash J and Bernier, Nathan R and Lukashchuk, Anton and Uhrich, Philipp and Qiu, Liu and Kippenberg, Tobias J},
  journal={Nature Electronics},
  volume={4},
  number={5},
  pages={326--332},
  month={May},
  year={2021},
  publisher={Nature Publishing Group UK London}
}

@inproceedings{de2021attojoule,
  title={Attojoule-per-bit electrical energy consumption optical modulators at {4 K} and {300 K} through energy harvesting},
  author={de Cea, Marc and Ram, Rajeev J},
  booktitle={IEEE Photonics Conference (IPC)},
  pages={1--2},
  month={October},
  year={2021}
}

@article{pintus2024cryogenic,
  title={Cryogenic optical data link for superconducting circuits},
  author={Pintus, Paolo and Soltani, Mo and Moody, Galan},
  journal={Nature Photonics},
  volume={18},
  number={4},
  pages={306--308},
  month={April},
  year={2024},
  publisher={Nature Publishing Group UK London}
}

@inproceedings{yin2021electronic,
  title={Electronic-photonic cryogenic egress link},
  author={Yin, Bozhi and Gevorgyan, Hayk and Onural, Deniz and Khilo, Anatol and Popovi{\'c}, Milo{\v{s}} A and Stojanovi{\'c}, Vladimir M},
  booktitle={IEEE 47th European Solid State Circuits Conference (ESSCIRC)},
  pages={51--54},
  month={September},
  year={2021}
}

@article{shen2024photonic,
  title={Photonic link from single-flux-quantum circuits to room temperature},
  author={Shen, Mohan and Xie, Jiacheng and Xu, Yuntao and Wang, Sihao and Cheng, Risheng and Fu, Wei and Zhou, Yiyu and Tang, Hong X},
  journal={Nature Photonics},
  volume={18},
  number={4},
  pages={371--378},
  month={April},
  year={2024},
  publisher={Nature Publishing Group UK London}
}

@article{krinner2019engineering,
  title={Engineering cryogenic setups for 100-qubit scale superconducting circuit systems},
  author={Krinner, Sebastian and Storz, Simon and Kurpiers, Philipp and Magnard, Paul and Heinsoo, Johannes and Keller, Raphael and Luetolf, Janis and Eichler, Christopher and Wallraff, Andreas},
  journal={EPJ Quantum Technology},
  volume={6},
  number={2},
  pages={1--29},
  month={May},
  year={2019},
  publisher={Springer Berlin Heidelberg}
}

@inproceedings{klostermann1991heat,
  title={Heat sinking of cryogenic coaxial cables in a dilution refrigerator},
  author={Klostermann, L and Trentalange, S and Ritzi, B and Wen, HC and Niinikoski, TO},
  booktitle={High Energy Spin Physics: Volume 2: Workshops},
  pages={378--384},
  year={1991},
  organization={Springer}
}

@article{robertazzi1997flux,
  title={Flux trapping experiments in single flux quantum shift registers},
  author={Robertazzi, RP and Siddiqi, I and Mukhanov, O},
  journal={IEEE Transactions on Applied Superconductivity},
  volume={7},
  number={2},
  pages={3164--3167},
  month={June},
  year={1997}
}

@article{fourie2021experimental,
  title={Experimental verification of moat design and flux trapping analysis},
  author={Fourie, Coenrad J and Jackman, Kyle},
  journal={IEEE Transactions on Applied Superconductivity},
  volume={31},
  number={5},
  note = {{A}rt. no. 1300507},
  month={August},
  year={2021}
}

@article{peng2019solution,
  title={A Solution for ultra-low bit-error-rate Interface of Superconductor-semiconductor by using an error-correction-code encoder},
  author={Peng, Xizhu and Liu, Xiaoqiao and Mei, Yajun and Ren, Jie and Tang, He},
  journal={IEEE Transactions on Applied Superconductivity},
  volume={29},
  number={5},
  note = {{A}rt. no. 1301604},
  month={August},
  year={2019}
}

@article{mustafa2025lightweight,
  title={Lightweight Error-Correction Code Encoders in Superconducting Electronic Systems},
  author={Mustafa, Yerzhan and Pek{\"o}z, Berker and K{\"o}se, Sel{\c{c}}uk},
  journal={arXiv preprint arXiv:2509.00962},
  year={2025}
}

@inproceedings{nagaoka2022microprocessor,
  title={A 57.2 {GHz} 11.2 {mW} 8-bit general purpose superconductor microprocessor with dual-clocking scheme},
  author={Nagaoka, Ikki and Kashima, Ryota and Nakano, Tomoki and Tanaka, Masamitsu and Yamashita, Taro and Inoue, Koji and Fujimaki, Akira},
  booktitle={IEEE Asian Solid-State Circuits Conference (A-SSCC)},
  pages={1--3},
  month={November},
  year={2022}
}

@article{kashima2021microprocessor,
  title={64-{GHz} datapath demonstration for bit-parallel {SFQ} microprocessors based on a gate-level-pipeline structure},
  author={Kashima, Ryota and Nagaoka, Ikki and Tanaka, Masamitsu and Yamashita, Taro and Fujimaki, Akira},
  journal={IEEE Transactions on Applied Superconductivity},
  volume={31},
  number={5},
  month={August},
  year={2021}
}

@article{tanaka2023execution,
  title={Execution of stored programs by a rapid single-flux-quantum random-access-memory-embedded bit-serial microprocessor using 50-{GHz} clock frequency},
  author={Tanaka, Masamitsu and Sato, Ryo and Fujimaki, Akira and Takagi, Kazuyoshi and Takagi, Naofumi},
  journal={Applied Physics Letters},
  volume={122},
  number={19},
  note = {{A}rt. no. 192601},
  month={May},
  year={2023},
  publisher={AIP Publishing}
}

@article{ando2016design,
  title={Design and demonstration of an 8-bit bit-serial {RSFQ} microprocessor: {CORE} e4},
  author={Ando, Yuki and Sato, Ryo and Tanaka, Masamitsu and Takagi, Kazuyoshi and Takagi, Naofumi and Fujimaki, Akira},
  journal={IEEE Transactions on Applied Superconductivity},
  volume={26},
  number={5},
  note = {{A}rt. no. 1301205},
  month={August},
  year={2016}
}

@article{ishida2018towards,
  title={Towards ultra-high-speed cryogenic single-flux-quantum computing},
  author={Ishida, Koki and Tanaka, Masamitsu and Ono, Takatsugu and Inoue, Koji},
  journal={IEICE Transactions on Electronics},
  volume={101},
  number={5},
  pages={359--369},
  month={May},
  year={2018},
  publisher={The Institute of Electronics, Information and Communication Engineers}
}

@article{alam2023review,
  title={Cryogenic memory technologies},
  author={Alam, Shamiul and Hossain, Md Shafayat and Srinivasa, Srivatsa Rangachar and Aziz, Ahmedullah},
  journal={Nature Electronics},
  volume={6},
  number={3},
  pages={185--198},
  month={March},
  year={2023}
}

@article{herr2024data,
  title={A data center in a shoebox: {Imec's} plan to use superconductors to shrink computers},
  author={Herr, Anna and Herr, Quentin},
  journal={IEEE Spectrum},
  volume={61},
  number={6},
  pages={37--41},
  month={June},
  year={2024}
}

@misc{snowcap_compute, 
title={{Reuters: Snowcap Compute raises \$23 million for superconducting AI chips}}, 
url={https://www.reuters.com/business/snowcap-compute-raises-23-million-superconducting-ai-chips-2025-06-23/}, 
publisher={Reuters},
month={June},
  year={2025}}

@misc{seeqc_nvidia, 
title={{Quantum Insider: SEEQC Develops Digital Interface for Real-Time Quantum-Classical Integration with NVIDIA-Powered Error Correction}}, 
url={https://thequantuminsider.com/2025/03/20/seeqc-develops-digital-interface-for-real-time-quantum-classical-integration-with-nvidia-powered-error-correction/}, 
publisher={Quantum Insider},
month={March},
  year={2025}}

@misc{seeqc_nvidia2, 
title={{SEEQC Announces Digital Chip-Based Collaboration with NVIDIA to Accelerate Quantum Supercomputing
}}, 
url={https://seeqc.com/seeqc-nvidia}, 
publisher={SEEQC},
note={{A}ccessed: Aug 14, 2025}}

@article{wikborg1999rsfq,
  title={{RSFQ} front-end for a software radio receiver},
  author={Wikborg, Erland B and Semenov, Vasili K and Likharev, Konstantin K},
  journal={IEEE Transactions on Applied Superconductivity},
  volume={9},
  number={2},
  pages={3615--3618},
  month={June},
  year={1999}
}

@article{kirichenko2005superconductor,
  title={Superconductor digital receiver components},
  author={Kirichenko, Alex and Sarwana, Saad and Gupta, Deepnarayan and Yohannes, Daniel},
  journal={IEEE Transactions on Applied Superconductivity},
  volume={15},
  number={2},
  pages={249--254},
  month={June},
  year={2005}
}

@article{gupta2007digital,
  title={Digital channelizing radio frequency receiver},
  author={Gupta, Deepnarayan and Filippov, Timur V and Kirichenko, Alexander F and Kirichenko, Dmitri E and Vernik, Igor V and Sahu, Anubhav and Sarwana, Saad and Shevchenko, Pavel and Talalaevskii, Andrei and Mukhanov, Oleg A},
  journal={IEEE Transactions on applied superconductivity},
  volume={17},
  number={2},
  pages={430--437},
  month={June},
  year={2007}
}

@article{mukhanov2008superconductor,
  title={Superconductor digital-{RF} receiver systems},
  author={Mukhanov, Oleg A and Kirichenko, Dmitri and Vernik, Igor V and Filippov, Timur V and Kirichenko, Alexander and Webber, Robert and Dotsenko, Vladimir and Talalaevskii, Andrei and Tang, Jia Cao and Sahu, Anubhav and others},
  journal={IEICE Transactions on Electronics},
  volume={91},
  number={3},
  pages={306--317},
  month={March},
  year={2008},
  publisher={The Institute of Electronics, Information and Communication Engineers}
}

@article{mitola2000software,
  title={Software radio architecture evolution: Foundations, technology tradeoffs, and architecture implications},
  author={MITOLA III, Joseph},
  journal={IEICE Transactions on Communications},
  volume={83},
  number={6},
  pages={1165--1173},
  month={June},
  year={2000},
  publisher={The Institute of Electronics, Information and Communication Engineers}
}

@misc{hypres_RF, 
title={{Advanced Digital-RF Receiver (ADR)}}, 
url={https://www.hypres.com/products/advanced-digital-rf-receiver/}, 
publisher={HYPRES},
note={{A}ccessed: Oct 23, 2025}}

@article{schneider2022supermind,
  title={{SuperMind}: a survey of the potential of superconducting electronics for neuromorphic computing},
  author={Schneider, Michael and Toomey, Emily and Rowlands, Graham and Shainline, Jeff and Tschirhart, Paul and Segall, Ken},
  journal={Superconductor Science and Technology},
  volume={35},
  number={5},
  note = {{A}rt. no. 053001},
  month={March},
  year={2022},
  publisher={IOP Publishing}
}

@article{islam2023review,
  title={A review of cryogenic neuromorphic hardware},
  author={Islam, Md Mazharul and Alam, Shamiul and Hossain, Md Shafayat and Roy, Kaushik and Aziz, Ahmedullah},
  journal={Journal of Applied Physics},
  volume={133},
  number={7},
  note = {{A}rt. no. 070701},
  month={February},
  year={2023},
  publisher={AIP Publishing}
}

@article{karamuftuoglu2023jj,
  title={{JJ-Soma}: Toward a spiking neuromorphic processor architecture},
  author={Karamuftuoglu, Mustafa Altay and Bozbey, Ali and Razmkhah, Sasan},
  journal={IEEE Transactions on Applied Superconductivity},
  volume={33},
  number={8},
  note = {{A}rt. no. 1400607},
  month={November},
  year={2023}
}

@article{razmkhah2024hybrid,
  title={Hybrid synaptic structure for spiking neural network realization},
  author={Razmkhah, Sasan and Karamuftuoglu, Mustafa Altay and Bozbey, Ali},
  journal={Superconductor Science and Technology},
  volume={37},
  number={6},
  note = {{A}rt. no. 065011},
  month={May},
  year={2024},
  publisher={IOP Publishing}
}

@article{przybysz2015superconductor,
  title={Superconductor digital electronics},
  author={Przybysz, John X and Miller, Donald L and Toepfer, Hannes and Mukhanov, Oleg and Lisenfeld, J{\"u}rgen and Weides, Martin and Rotzinger, Hannes and Febvre, Pascal},
  journal={Applied Superconductivity: Handbook on Devices and Applications},
  pages={1111--1206},
  year={2015},
  publisher={Wiley Online Library}
}

@article{braginski2019superconductor,
  title={Superconductor electronics: Status and outlook},
  author={Braginski, Alex I},
  journal={Journal of Superconductivity and Novel Magnetism},
  volume={32},
  number={1},
  pages={23--44},
  month={November},
  year={2019},
  publisher={Springer}
}

@article{holmes2013energy,
  title={Energy-efficient superconducting computing—Power budgets and requirements},
  author={Holmes, D Scott and Ripple, Andrew L and Manheimer, Marc A},
  journal={IEEE Transactions on Applied Superconductivity},
  volume={23},
  number={3},
  note = {{A}rt. no. 1701610},
  month={June},
  year={2013}
}

@article{kirichenko2011zero,
  title={Zero static power dissipation biasing of {RSFQ} circuits},
  author={Kirichenko, DE and Sarwana, Saad and Kirichenko, AF},
  journal={IEEE Transactions on Applied Superconductivity},
  volume={21},
  number={3},
  pages={776--779},
  month={June},
  year={2011}
}

@article{mukhanov2011energy,
  title={Energy-efficient single flux quantum technology},
  author={Mukhanov, Oleg A},
  journal={IEEE Transactions on Applied Superconductivity},
  volume={21},
  number={3},
  pages={760--769},
  month={June},
  year={2011}
}

@article{herr2011ultra,
  title={Ultra-low-power superconductor logic},
  author={Herr, Quentin P and Herr, Anna Y and Oberg, Oliver T and Ioannidis, Alexander G},
  journal={Journal of Applied Physics},
  volume={109},
  number={10},
  note = {{A}rt. no. 103903},
  month={May},
  year={2011},
  publisher={AIP Publishing}
}

@article{takeuchi2013adiabatic,
  title={An adiabatic quantum flux parametron as an ultra-low-power logic device},
  author={Takeuchi, Naoki and Ozawa, Dan and Yamanashi, Yuki and Yoshikawa, Nobuyuki},
  journal={Superconductor Science and Technology},
  volume={26},
  number={3},
  note = {{A}rt. no. 035010},
  month={January},
  year={2013},
  publisher={IOP Publishing}
}

@article{herr2023superconducting,
  title={Superconducting pulse conserving logic and {Josephson-SRAM}},
  author={Herr, Quentin and Josephsen, Trent and Herr, Anna},
  journal={Applied Physics Letters},
  volume={122},
  number={18},
  note = {{A}rt. no. 182604},
  month={May},
  year={2023},
  publisher={AIP Publishing}
}

@misc{lupo2025digital,
  title={Digital phase source for {Josephson} junction computing},
  author={Lupo, Federico Vittorio and Mukhanov, Oleg A and Arzeo, Marco},
  year={2025},
  month={March},
  publisher={Google Patents},
  note={{US} Patent App. 18/822,229}
}

@article{ayala2021mana,
  title={{MANA}: A monolithic adiabatic integration architecture microprocessor using 1.4-{zJ}/op unshunted superconductor {Josephson} junction devices},
  author={Ayala, Christopher L and Tanaka, Tomoyuki and Saito, Ro and Nozoe, Mai and Takeuchi, Naoki and Yoshikawa, Nobuyuki},
  journal={IEEE Journal of Solid-State Circuits},
  volume={56},
  number={4},
  pages={1152--1165},
  month={April},
  year={2021}
}

@inproceedings{kundu2025system,
  title={A System Level Performance Evaluation for Superconducting Digital Systems},
  author={Kundu, Joyjit and Bhattacharjee, Debjyoti and Josephsen, Nathan and Pokhrel, Ankit and De Silva, Udara and Guo, Wenzhe and Van Winckel, Steven and Brebels, Steven and Herr, Quentin and Herr, Anna and others},
  booktitle={Design, Automation \& Test in Europe Conference (DATE)},
  pages={1--7},
  month={March},
  year={2025}
}

@article{hironaka2020demonstration,
  title={Demonstration of a single-flux-quantum microprocessor operating with {Josephson-CMOS} hybrid memory},
  author={Hironaka, Yuki and Yamanashi, Yuki and Yoshikawa, Nobuyuki},
  journal={IEEE Transactions on Applied Superconductivity},
  volume={30},
  number={7},
  note={{A}rt. no. 1301206},
  month={October},
  year={2020}
}

@article{nagasawa1995380,
  title={A 380 ps, 9.5 {mW} {Josephson} {4-Kbit} {RAM} operated at a high bit yield},
  author={Nagasawa, Shuichi and Hashimoto, Yoshihito and Numata, Hideaki and Tahara, Shuichi},
  journal={IEEE Transactions on Applied Superconductivity},
  volume={5},
  number={2},
  pages={2447--2452},
  month={June},
  year={1995}
}

@article{nagasawa2007yield,
  title={Yield Evaluation of {10-kA/cm$^2$}  {Nb} Multi-Layer Fabrication Process Using Conventional Superconducting {RAMs}},
  author={Nagasawa, Shuichi and Satoh, Tetsuro and Hinode, Kenji and Kitagawa, Yoshihiro and Hidaka, Mutsuo},
  journal={IEEE Transactions on Applied Superconductivity},
  volume={17},
  number={2},
  pages={177--180},
  month={June},
  year={2007}
}

@inproceedings{mukhanov2019scalable,
  title={Scalable quantum computing infrastructure based on superconducting electronics},
  author={Mukhanov, O and Kirichenko, A and Howington, C and Walter, J and Hutchings, M and Vernik, I and Yohannes, D and Dodge, K and Ballard, A and Plourde, BLT and others},
  booktitle={IEEE International Electron Devices Meeting (IEDM)},
  pages={31.2.1--31.2.4},
  month={December},
  year={2019},
  organization={IEEE}
}

@inproceedings{jokar2022digiq,
  title={{DigiQ}: A scalable digital controller for quantum computers using {SFQ} logic},
  author={Jokar, Mohammad Reza and Rines, Richard and Pasandi, Ghasem and Cong, Haolin and Holmes, Adam and Shi, Yunong and Pedram, Massoud and Chong, Frederic T},
  booktitle={IEEE International Symposium on High-Performance Computer Architecture (HPCA)},
  pages={400--414},
  month={April},
  year={2022},
  organization={IEEE}
}

@article{mcdermott2018quantum,
  title={Quantum--classical interface based on single flux quantum digital logic},
  author={McDermott, Robert and Vavilov, Mikhail Grigorievich and Plourde, Benjamin Louis Thomas and Wilhelm, Frank Karl and Liebermann, Peter James and Mukhanov, Oleg Alexandrovich and Ohki, Tatsuya Akira},
  journal={Quantum science and technology},
  volume={3},
  number={2},
  note = {{A}rt. no. 024004},
  month={January},
  year={2018},
  publisher={IOP Publishing}
}

@article{mcdermott2014accurate,
  title={Accurate qubit control with single flux quantum pulses},
  author={McDermott, R and Vavilov, MG},
  journal={Physical Review Applied},
  volume={2},
  number={1},
  note = {{A}rt. no. 014007},
  month={July},
  year={2014},
  publisher={APS}
}

@article{patel2017phonon,
  title={Phonon-mediated quasiparticle poisoning of superconducting microwave resonators},
  author={Patel, U and Pechenezhskiy, Ivan V and Plourde, BLT and Vavilov, MG and McDermott, R},
  journal={Physical Review B},
  volume={96},
  number={22},
  note = {{A}rt. no. 220501},
  month={December},
  year={2017},
  publisher={APS}
}

@article{di2023discriminating,
  title={Discriminating the phase of a coherent tone with a flux-switchable superconducting circuit},
  author={Di Palma, Luigi and Miano, Alessandro and Mastrovito, Pasquale and Massarotti, Davide and Arzeo, Marco and Pepe, Giovanni Piero and Tafuri, Francesco and Mukhanov, O},
  journal={Physical Review Applied},
  volume={19},
  number={6},
  note={{A}rt. no. 064025},
  month={June},
  year={2023},
  publisher={APS}
}

@article{di2024fast,
  title={Fast Digital Phase Detection of a Coherent Tone At {GHz} Frequencies},
  author={Di Palma, L and Mastrovito, P and Miano, A and Salim, AJ and Lupo, FV and Bernhardt, J and Di Marino, L and Massarotti, D and Pepe, GP and Tafuri, F and others},
  journal={IEEE Transactions on Applied Superconductivity},
  volume={34},
  number={3},
  note={{A}rt. no. 1300305},
  month={May},
  year={2024}
}

@article{di2025control,
  title={Control of a {Josephson} digital phase detector via an {SFQ}-based flux bias driver},
  author={Di Marino, Laura and Di Palma, Luigi and Riccio, Michele and Arzeo, Marco and Mukhanov, Oleg},
  journal={IEEE Transactions on Quantum Engineering},
  volume={6},
  note={{A}rt. no. 3101708},
  month={June},
  year={2025}
}

@article{bernhardt2025quantum,
  title={Quantum Computer Controlled by Superconducting Digital Electronics at Millikelvin Temperature},
  author={Bernhardt, Jacob and Jordan, Caleb and Rahamim, Joseph and Kirchenko, Alex and Bharadwaj, Karthik and Fry-Bouriaux, Louis and Porsch, Katie and Somoroff, Aaron and Tsai, Kan-Ting and Walter, Jason and others},
  journal={arXiv preprint arXiv:2503.09879},
  month={March},
  year={2025}
}

@article{barbosa2024rsfq,
  title={{RSFQ} All-Digital Programmable Multi-Tone Generator For Quantum Applications},
  author={Barbosa, Jo{\~a}o and Brennan, Jack C and Casaburi, Alessandro and Hutchings, MD and Kirichenko, Alex and Mukhanov, Oleg and Weides, Martin},
  journal={IEEE Transactions on Quantum Engineering},
  volume={6},
  note={{A}rt. no. 5500211},
  month={December},
  year={2024}
}

@misc{kannan2024managing,
  title={Managing coupling in a quantum computing system},
  author={KANNAN, Bharath and Sung, Youngkyu and Ding, Leon Chen and Menke, Tim and Novikov, Sergey Sergeevich and Gustavsson, Simon Karl Fredrik and Oliver, William David},
  year={2024},
  month={April},
  publisher={Google Patents},
  note={{US} Patent App. 18/477,603}
}

@article{liu2023single,
  title={Single flux quantum-based digital control of superconducting qubits in a multichip module},
  author={Liu, Chuan-Hong and Ballard, Andrew and Olaya, David and Schmidt, Daniel R and Biesecker, John and Lucas, Tammy and Ullom, Joel and Patel, Shravan and Rafferty, Owen and Opremcak, Alexander and others},
  journal={PRX Quantum},
  volume={4},
  number={3},
  note={{A}rt. no. 030310},
  month={July},
  year={2023},
  publisher={APS}
}

@misc{google_quantum_ai, 
title={{We’re scaling quantum computing even faster with Atlantic Quantum.
}}, 
url={https://blog.google/technology/research/scaling-quantum-computing-even-faster-with-atlantic-quantum/}, 
publisher={Google},
note={{Oct.} 2, 2025}}

@inproceedings{komissarov2024modified,
  title={Modified Superconducting Single-Flux Quantum Two-Photon Coincidence Correlator for Single-Photon Measurements},
  author={Komissarov, Ivan V and Salim, Amir J and Mukhanov, Oleg A and Rambo, Timothy and Miller, Aaron and Sobolewski, Roman},
  booktitle={CLEO: Applications and Technology},
  pages={JW2A--89},
  month={May},
  year={2024},
  organization={Optica Publishing Group}
}

@inproceedings{sobolewski2025electrical,
  title={Electrical testing of {SNSPD-SFQ} two-photon coincidence correlator},
  author={Sobolewski, Roman and Komissarov, Ivan and Salim, Amir J and Walter, Jason and Yohannes, Daniel and Mukhanov, Oleg and Talalaevski, Andrei and Track, Elie and Rambo, Tim and Miller, Aaron},
  booktitle={Quantum Optics and Photon Counting},
  pages={PC1352506},
  month={June},
  year={2025},
  organization={SPIE}
}

@article{gol2001picosecond,
  title={Picosecond superconducting single-photon optical detector},
  author={Gol’Tsman, GN and Okunev, O and Chulkova, G and Lipatov, A and Semenov, A and Smirnov, K and Voronov, B and Dzardanov, A and Williams, C and Sobolewski, Roman},
  journal={Applied Physics Letters},
  volume={79},
  number={6},
  pages={705--707},
  month={August},
  year={2001},
  publisher={American Institute of Physics}
}

@article{miyajima2017timing,
  title={Timing discriminator based on single-flux-quantum circuit toward high time-resolved photon detection},
  author={Miyajima, Shigeyuki and Miki, Shigehito and Yabuno, Masahiro and Yamashita, Taro and Terai, Hirotaka},
  journal={Superconductor Science and Technology},
  volume={30},
  number={12},
  pages={12LT01},
  month={October},
  year={2017},
  publisher={IOP Publishing}
}

@article{miki2018superconducting,
  title={Superconducting coincidence photon detector with short timing jitter},
  author={Miki, S and Miyajima, S and Yabuno, M and Yamashita, T and Yamamoto, T and Imoto, N and Ikuta, R and Kirkwood, RA and Hadfield, RH and Terai, H},
  journal={Applied Physics Letters},
  volume={112},
  number={26},
  note={{A}rt. no. 262601},
  month={June},
  year={2018},
  publisher={AIP Publishing}
}

@inproceedings{shelly2017modelling,
  title={Modelling of a two-signal {SFQ} detection scheme for the readout of superconducting nanowire single photon detectors},
  author={Shelly, Connor D and See, Patrick and Romans, Ed J and Casaburi, Alessandro and Ireland, Jane and Williams, Jonathan M and Hadfield, Robert H},
  booktitle={16th International Superconductive Electronics Conference (ISEC)},
  pages={1--3},
  month={June},
  year={2017}
}

@article{kiviranta2025two,
  title={Two-Stage {SQUID} Amplifier With Bias Current Re-Use},
  author={Kiviranta, Mikko and Gr{\"o}nberg, Leif},
  journal={IEEE Transactions on Applied Superconductivity},
  volume={35},
  number={5},
  note={{A}rt. no. 1600304},
  month={August},
  year={2025}
}

@article{hummatov2023fast,
  title={Fast transition-edge sensors suitable for photonic quantum computing},
  author={Hummatov, Ruslan and Lita, Adriana E and Farrahi, Tannaz and Otrooshi, Negar and Fayer, Samuel and Collins, Matthew J and Durkin, Malcolm and Bennett, Douglas and Ullom, Joel and Mirin, Richard P and others},
  journal={Journal of Applied Physics},
  volume={133},
  number={23},
  note={{A}rt. no. 234502},
  month={June},
  year={2023},
  publisher={AIP Publishing}
}

@article{gottardi2021review,
  title={A review of {X}-ray microcalorimeters based on superconducting transition edge sensors for astrophysics and particle physics},
  author={Gottardi, Luciano and Nagayashi, Kenichiro},
  journal={Applied Sciences},
  volume={11},
  number={9},
  note={{A}rt. no. 3793},
  month={April},
  year={2021},
  publisher={MDPI}
}

@article{ullom2015review,
  title={Review of superconducting transition-edge sensors for x-ray and gamma-ray spectroscopy},
  author={Ullom, Joel N and Bennett, Douglas A},
  journal={Superconductor Science and Technology},
  volume={28},
  number={8},
  note={{A}rt. no. 084003},
  month={July},
  year={2015},
  publisher={IOP Publishing}
}

@article{beyer2003performance,
  title={Performance of 32-channel time-division {SQUID} multiplexer for cryogenic detector arrays},
  author={Beyer, Joern and De Korte, PAJ and Reintsema, Carl D and Nam, Sae Woo and MacIntosh, M and Hilton, GC and Vale, LR and Irwin, KD},
  journal={IEEE Transactions on Applied Superconductivity},
  volume={13},
  number={2},
  pages={649--652},
  month={June},
  year={2003}
}

@article{kiviranta2021two,
  title={Two-stage {SQUID} amplifier for the frequency multiplexed readout of the {X-IFU X}-ray camera},
  author={Kiviranta, Mikko and Gr{\"o}nberg, Leif and Puranen, Tuomas and Van der Kuur, Jan and Beev, Nikolai and Salonen, Jaakko and Hazra, Dibyendu and Korpela, Seppo},
  journal={IEEE Transactions on Applied Superconductivity},
  volume={31},
  number={5},
  note={{A}rt. no. 1600605},
  month={August},
  year={2021}
}

@article{durkin2019demonstration,
  title={Demonstration of {Athena X-IFU} compatible 40-row time-division-multiplexed readout},
  author={Durkin, Malcolm and Adams, Joseph S and Bandler, Simon R and Chervenak, James A and Chaudhuri, Saptarshi and Dawson, Carl S and Denison, Edward V and Doriese, William B and Duff, Shannon M and Finkbeiner, Fred M and others},
  journal={IEEE Transactions on Applied Superconductivity},
  volume={29},
  number={5},
  note={{A}rt. no. 2101005},
  month={August},
  year={2019}
}

@article{bozbey2009single,
  title={Single-flux-quantum circuit based readout system for detector arrays by using time to digital conversion},
  author={Bozbey, Ali and Miyajima, Shigeyuki and Akaike, Hiroyuki and Fujimaki, Akira},
  journal={IEEE Transactions on Applied Superconductivity},
  volume={19},
  number={3},
  pages={509--513},
  month={June},
  year={2009}
}

@article{leman2023integrated,
  title={Integrated Superconducting Transition-Edge-Sensor Energy Readout ({ISTER})},
  author={Leman, Steven W and Golden, Evan B and Guyton, Matthew C and Ryu, Kevin K and Wynn, Alex},
  journal={IEEE Transactions on Applied Superconductivity},
  volume={33},
  number={5},
  note={{A}rt. no. 2500807},
  month={August},
  year={2023}
}

\end{document}